\documentclass[twocolumn]{aastex62}
\usepackage{CJK}

\accepted{\today}
\submitjournal{ApJ}

\shorttitle{The Magnetic Field in $\rho$ Oph C}
\shortauthors{Liu et al.}

\begin{document}
\title{The JCMT BISTRO Survey: The Magnetic Field In The Starless Core $\rho$ Ophiuchus C}

%me and my supervisor

\correspondingauthor{Keping Qiu}
\email{kpqiu@nju.edu.cn}

\author[0000-0002-4774-2998]{Junhao Liu}
\affil{School of Astronomy and Space Science, Nanjing University, 163 Xianlin Avenue, Nanjing 210023, People's Republic of China}
\affil{Key Laboratory of Modern Astronomy and Astrophysics (Nanjing University), Ministry of Education, Nanjing 210023, People's Republic of China}
\affil{Harvard-Smithsonian Center for Astrophysics, 60 Garden Street, Cambridge, MA 02138, USA}

\author[0000-0002-5093-5088]{Keping Qiu}
\affil{School of Astronomy and Space Science, Nanjing University, 163 Xianlin Avenue, Nanjing 210023, People's Republic of China}
\affil{Key Laboratory of Modern Astronomy and Astrophysics (Nanjing University), Ministry of Education, Nanjing 210023, People's Republic of China}

%data reduction is important, especially for Oph-C...

\author{David Berry}
\affil{East Asian Observatory, 660 N. A`oh\={o}k\={u} Place, University Park, Hilo, HI 96720, USA}

%who gave me comments

\author[0000-0002-9289-2450]{James Di Francesco}
\affil{NRC Herzberg Astronomy and Astrophysics, 5071 West Saanich Road, Victoria, BC V9E 2E7, Canada}
\affil{Department of Physics and Astronomy, University of Victoria, Victoria, BC V8P 1A1, Canada}

\author[0000-0002-0794-3859]{Pierre Bastien} 
\affiliation{Institut de Recherche sur les Exoplan\`etes (iREx), Universit\'e de Montr\'eal, D\'epartement de Physique, C.P. 6128 Succ. Centre-ville, Montr\'eal, QC, H3C 3J7, Canada}
\affiliation{Centre de Recherche en Astrophysique du Qu\'ebec (CRAQ), Universit\'e de Montr\'eal, D\'epartement de Physique, C.P. 6128 Succ. Centre-ville, Montr\'eal, QC, H3C 3J7, Canada}

\author[0000-0003-2777-5861]{Patrick M. Koch}
\affil{Academia Sinica Institute of Astronomy and Astrophysics, P.O. Box 23-141, Taipei 10617, Taiwan}

\author[0000-0003-0646-8782]{Ray S. Furuya}
\affil{Tokushima University, Minami Jousanajima-machi 1-1, Tokushima 770-8502, Japan}
\affil{Institute of Liberal Arts and Sciences Tokushima University, Minami Jousanajima-machi 1-1, Tokushima 770-8502, Japan}

\author[0000-0003-2412-7092]{Kee-Tae Kim}
\affil{Korea Astronomy and Space Science Institute, 776 Daedeokdae-ro, Yuseong-gu, Daejeon 34055, Republic of Korea}

\author[0000-0002-0859-0805]{Simon Coud\'{e}}
\affil{SOFIA Science Center, Universities Space Research Association, NASA Ames Research Center, M.S. N232-12, Moffett Field, CA 94035, USA}
\affiliation{Centre de Recherche en Astrophysique du Qu\'ebec (CRAQ), Universit\'e de Montr\'eal, D\'epartement de Physique, C.P. 6128 Succ. Centre-ville, Montr\'eal, QC, H3C 3J7, Canada}

\author[0000-0002-3179-6334]{Chang Won Lee}
\affil{Korea Astronomy and Space Science Institute, 776 Daedeokdae-ro, Yuseong-gu, Daejeon 34055, Republic of Korea}
\affil{Korea University of Science and Technology, 217 Gajang-ro, Yuseong-gu, Daejeon 34113, Republic of Korea}

% ...

%\author{Joe Mottram}
%\affil{Max-Planck-Institut f{\"u}r Radioastronomie, Auf dem H{\"u}gel 69, D-53121 Bonn, Germany}

\begin{abstract}
We report 850~$\mu$m dust polarization observations of a low-mass ($\sim$12 $M_{\odot}$) starless core in the $\rho$ Ophiuchus cloud, Ophiuchus C, made with the POL-2 instrument on the James Clerk Maxwell Telescope (JCMT) as part of the JCMT B-fields In STar-forming Region Observations (BISTRO) survey. We detect an ordered magnetic field projected on the plane of sky in the starless core. The magnetic field across the $\sim$0.1~pc core shows a predominant northeast-southwest orientation centering between $\sim$40$\degr$ to $\sim$100$\degr$, indicating that the field in the core is well aligned with the magnetic field in lower-density regions of the cloud probed by near-infrared observations and also the cloud-scale magnetic field traced by Planck observations. The polarization percentage ($P$) decreases with an increasing total intensity ($I$) with a power-law index of $-$1.03 $\pm$ 0.05. We estimate the plane-of-sky field strength ($B_{\mathrm{pos}}$) using modified Davis-Chandrasekhar-Fermi (DCF) methods based on structure function (SF), auto-correlation (ACF), and unsharp masking (UM) analyses. We find that the estimates from the SF, ACF, and UM methods yield strengths of 103 $\pm$ 46 $\mu$G, 136 $\pm$ 69 $\mu$G, and 213 $\pm$ 115 $\mu$G, respectively. Our calculations suggest that the Ophiuchus C core is near magnetically critical or slightly magnetically supercritical (i.e. unstable to collapse). The total magnetic energy calculated from the SF method is comparable to the turbulent energy in Ophiuchus C, while the ACF method and the UM method only set upper limits for the total magnetic energy because of large uncertainties.  
\end{abstract}


\section{introduction}
The role of magnetic fields (B-fields) has long been a hot topic under debates in the star formation studies \citep{2012ARA&A..50...29C}. There are two major classes of star-formation theories that significantly differ in the role played by magnetic fields. ``Strong magnetic field models'' suggest that molecular clouds are supported by magnetic fields, which quasi-statically dissipate via ambipolar diffusion. Eventually self-gravity overcomes the magnetic force, inducing the collapse of molecular cloud cores and the formation of stars \citep{2006ApJ...646.1043M}. In contrast, ``weak field models'' suggest that turbulent flows, instead of magnetic fields, dominate the evolution of molecular clouds and create overdense regions where stars form \citep{2004RvMP...76..125M}. Recently, results from simulations indicate that magnetic field and turbulence are both essential to provide support against gravitational collapse \citep[][and references therein]{2014prpl.conf...77P}. Observational studies of magnetic fields in star-forming regions can directly test these theoretical models, providing deep insights into the relative importance of magnetic fields and gravity/turbulence in cloud evolution and star formation. 

Observing the polarized emission of dust grains and the polarization of background stars is one of powerful ways to investigate the plane-of-sky magnetic field structure in star-forming regions \citep{1988QJRAS..29..327H}. The starlight polarization was first discovered by \citet{1949Sci...109..165H} and \citet{1949Sci...109..166H}. Later on, the observed polarization of starlight was explained by the partial extinction of starlight by magnetically aligned dust grains \citep{1988ApL&C..26..263H}, where the short axes of spinning dust grains align with magnetic field lines. This explaination is widely accepted. There are many theories trying to explain why dust grains are aligned with magnetic fields. Among them, the Radiative Alighment Torque (RAT) theory is most accepted \citep{2007JQSRT.106..225L}. Although the detail of the the gain alignment mechanism is still unclear, the plane-of-sky magnetic field structure in star formation regions has been successfully traced using polarization observations \citep{2012ARA&A..50...29C}. Polarization observations at near-infrared (NIR) wavelengths, which are expected to trace polarization produced by dust extinction of background starlight, are often used to investigate the magnetic filed structure in dense molecular regions \citep{2014ApJ...783....1S, 2015ApJS..220...17K}. However, NIR polarization observations are not sufficient to trace the magnetic field in regions with high extinction or associated with few background stars. Polarization observations at sub-millimeter (sub-mm) wavelengths, which trace dust thermal emission, are essential to overcome the drawback of NIR polarization observations and to probe the magnetic field structure in denser enviroments such as filaments and dense cores.

Among an increasing number of polarization observations toward low-mass star formation regions, studies of the protostellar phase of Young Stellar Objects (YSO) have attracted most interests. One important approach of these studies is to find hourglass-shaped magnetic fields. As predicted by the theoretical model and simulations \citep{1993ApJ...417..220G, 1993ApJ...417..243G}, the magnetic field of magnetically dominated dense regions is expected to show an hourglass shape in the collapse phase. At 0.001-0.01 pc scales, dust polarization observations towards low-mass protostellar systems have revealed the expected hourglass-shaped field morphologies \citep{2006Sci...313..812G, 2009ApJ...707..921R, 2013ApJ...769L..15S}. More chaotic field morphologies, which are expected in weakly magnetic environment or probably affected by stellar feedback or complex geometry, are also reported at this scale \citep{2014ApJS..213...13H, 2017ApJ...842L...9H}. At larger scales (0.01-0.1 pc), hints of the hourglass shape are less obvious \citep{2009ApJS..182..143M, 2010ApJS..186..406D, 2014ApJS..213...13H} and the role of magnetic field at this scale is comparatively less understood.

Since the magnetic field in protostellar cores can suffer from feedback by star-forming activities, polarization studies of starless cores are essential to help us understand the role of magnetic field in the early stages of star formation. The relatively weak polarized dust emission in starless cores, however, is far more difficult to detect and, as a result, there are only a handful of dust polarization observations toward starless cores \citep{2000ApJ...537L.135W, 2004ApJ...600..279C, 2009MNRAS.398..394W, 2014A&A...569L...1A}. The role of magnetic fields in the initial phase of star formation remains an open question.  

The Ophiuchus molecular cloud is a low-mass star-forming region located at a distance of $\sim$137 pc \citep{2017ApJ...834..141O}. It is one of the nearest star formation regions and has been widely studied \citep[][and references therein]{2008hsf2.book..351W}. Star formation in this cloud is heavily influenced by compression of expanding shock shells from the nearby Sco-Cen OB association \citep{1977AJ.....82..198V}. A detailed DCO$^+$ emission study has identified several dense cores, Ophiuchus A to Ophiuchus F (hereafter Oph-A to Oph-F), in the main body of Ophiuchus \citep{1990ApJ...365..269L}. Among these dense cores, our target, Oph-C, which harbors no embedded protostars \citep{2009ApJ...692..973E} and is not associated with Herschel 70 $\mu$m emission \citep{2015MNRAS.450.1094P}, appears to be the least evolved and is extremely quiescent. The 850 $\mu$m continuum of Oph-C was observed as part of the James Clerk Maxwell Telescope (JCMT) Gould Belt Survey (GBS) \citep{2007PASP..119..855W, 2015MNRAS.450.1094P}. \citet{2015MNRAS.450.1094P} identified a few low-mass, pressure-confined, virially bound, and $\sim$0.01 pc-scale ($\sim$ 3000 AU) sub-cores in Oph-C based on the GBS data. The 850 $\mu$m polarization data of the Ophiuchus cloud obtained using SCUPOL, the previous JCMT polarimeter, was catalogued by \citet{2009ApJS..182..143M}. More recently, the large-scale plane-of-sky magnetic field map of the Ophiuchus cloud was presented by \citet{2015A&A...576A.105P} at 5$\arcmin$ resolution as part of the Planck project. \citet{2015ApJS..220...17K} conducted NIR polarimetry of Ophiuchus cloud and suggested the magnetic field structures in the cloud may have been influenced by the nearby Sco-Cen OB association.

Here we present 850 $\mu$m dust polarization observations using the POL-2 polarimeter in combination with the Summillimetre Common-User Bolometer Array 2 (SCUBA-2) on the JCMT toward the Oph-C region as part of the B-fields In STar-forming Region Observations (BISTRO) survey \citep{2017ApJ...842...66W}. The BISTRO survey is aimed at using POL-2 to map the polarized dust emission in the densest parts of all of the Gould Belt star-forming regions including Orion A \citep{2017ApJ...846..122P}, Oph-A \citep{2018ApJ...859....4K}, M16 \citep{2018ApJ...860L...6P}, Oph-B \citep{2018ApJ...861...65S}, and several other regions (papers in preparation). With the unique resolution offered by JCMT, which can resolve the magnetic field structures down to scales of $\sim$1000 AU in nearby star formation regions, these POL-2 observations are crucial to test theoretical models of star formation at an intermediate scale and to generate a large sample of polarization maps of dense cores obtained in a uniform and consistent way for statistical studies \citep{2017ApJ...842...66W}. The B-field structures traced by POL-2 agree well with those traced by the previous SCUPOL observations, but the POL-2 maps are more sensitive than the previous SCUPOL data and trace larger areas \citep[e.g.,][]{2018ApJ...859....4K, 2018ApJ...861...65S}.  

This paper is organized as follows: in section 2, we describe the observations and data reduction; in section 3, we present the results of the observations and derive the B-field strength; in section 4, we discuss our results; and section 5 is given for a summary of this paper..

\section{Observations}
\begin{figure*}[!tbp]
\gridline{\fig{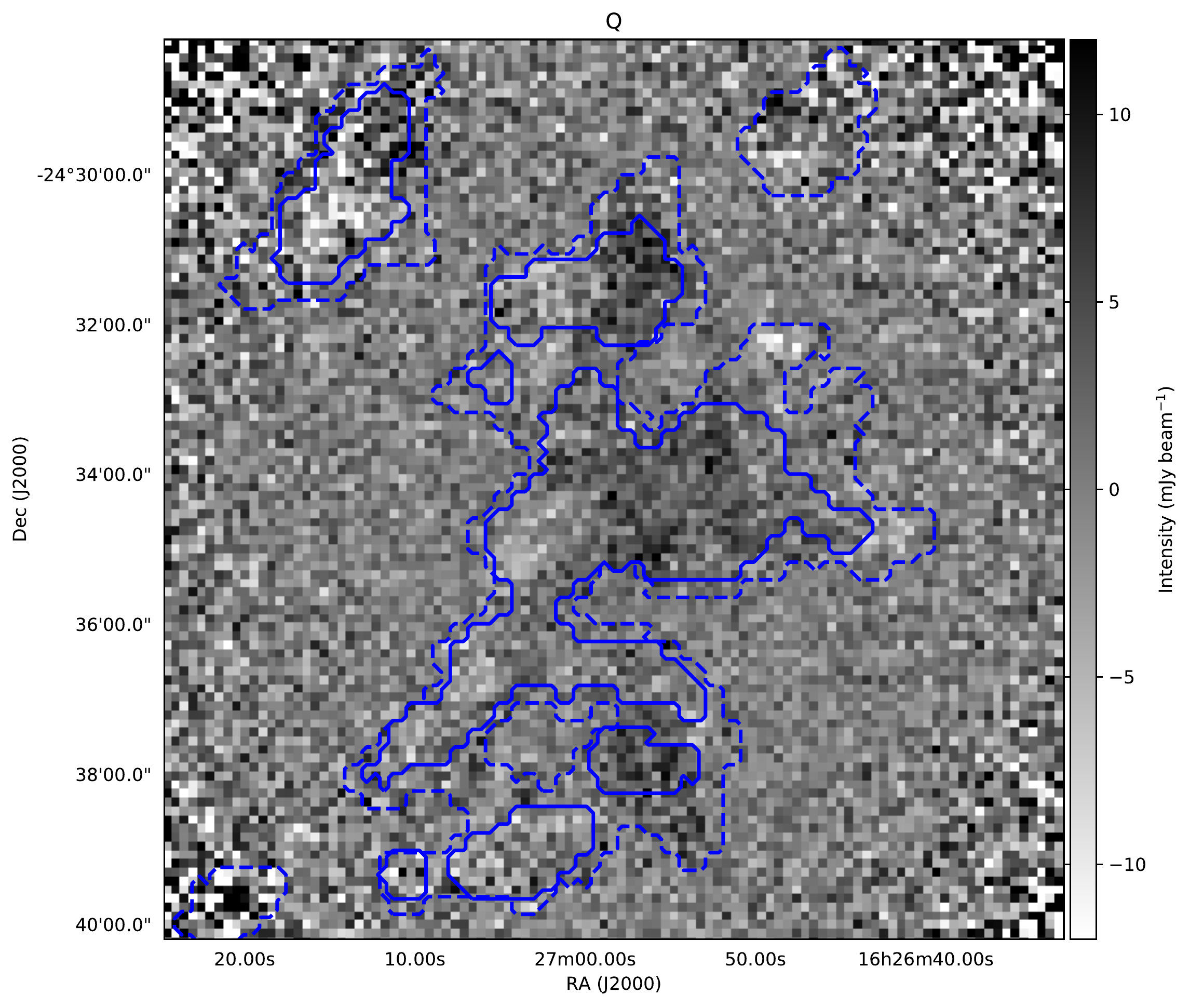}{0.5\textwidth}{(a)}
        \fig{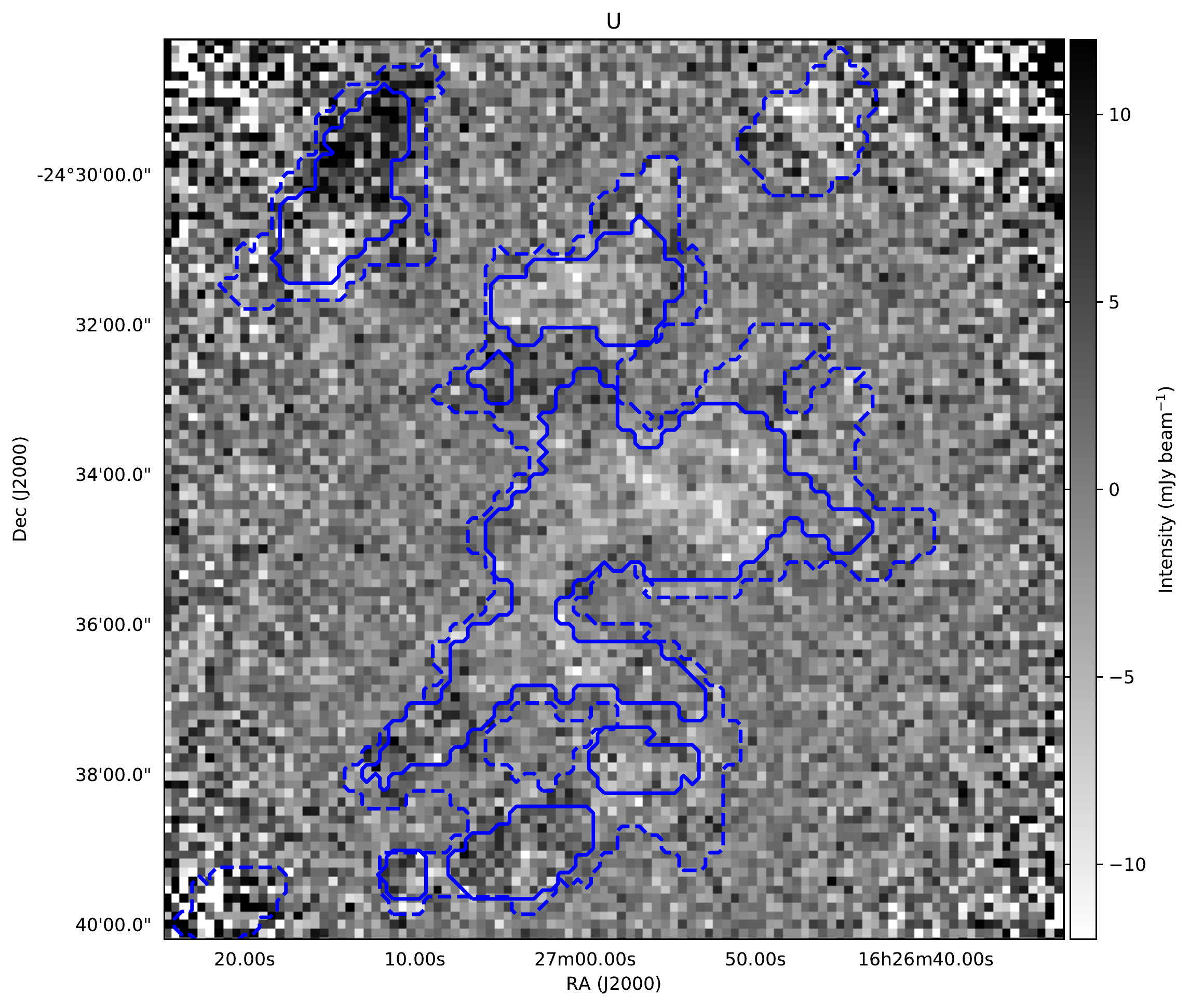}{0.5\textwidth}{(b)}}
 \gridline{\fig{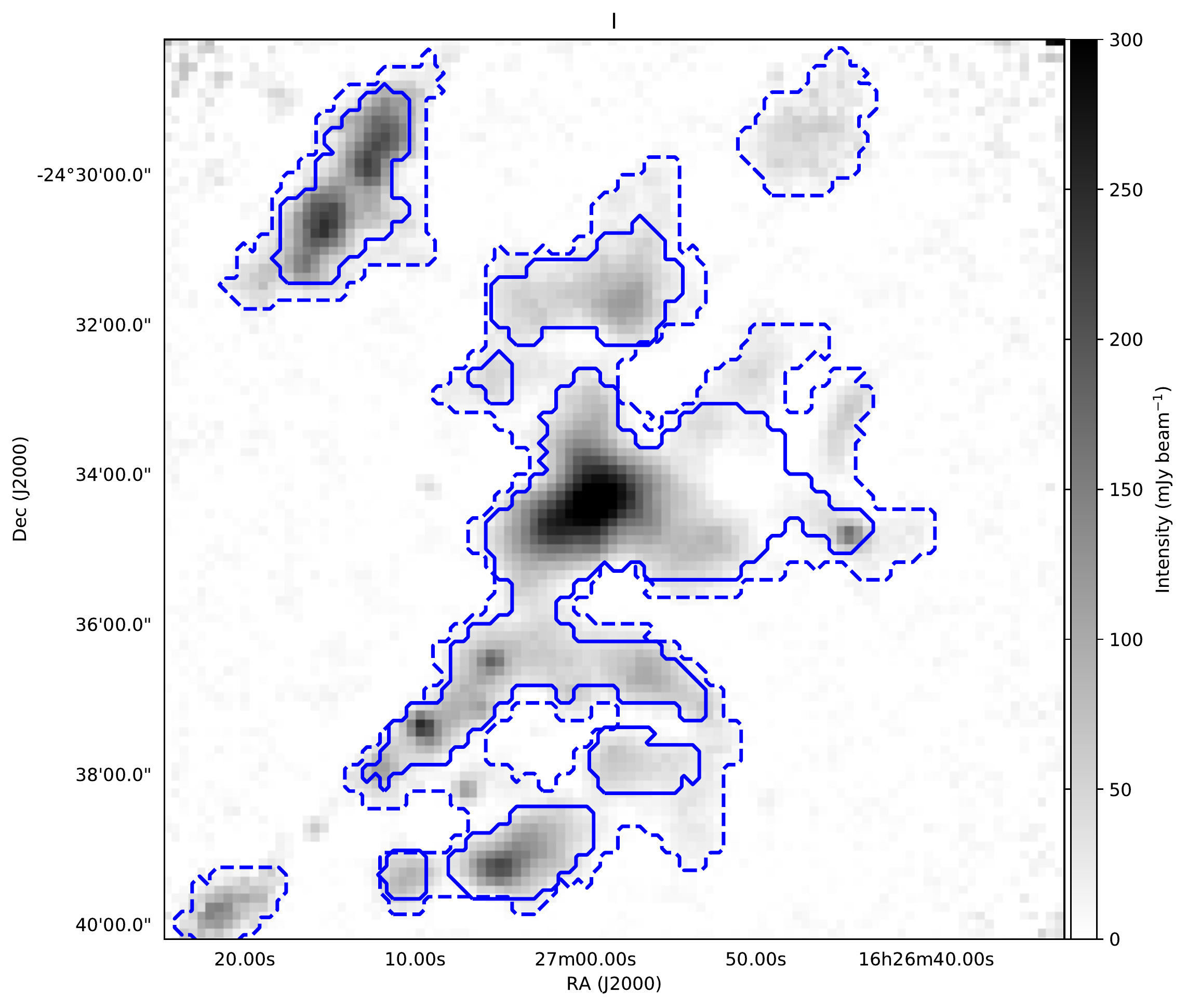}{0.5\textwidth}{(c)}}
\caption{(a)-(c) POL-2 Stokes $Q$, $U$, and $I$ maps of the Oph-C region. The intensity is shown in grey scale. The ASTMASK and the PCAMASK used in the data reduction process are shown in dashed line and solid line, respectively.  \label{fig:figiqu}}
\end{figure*}

\begin{figure*}[!tbp]

 \gridline{\fig{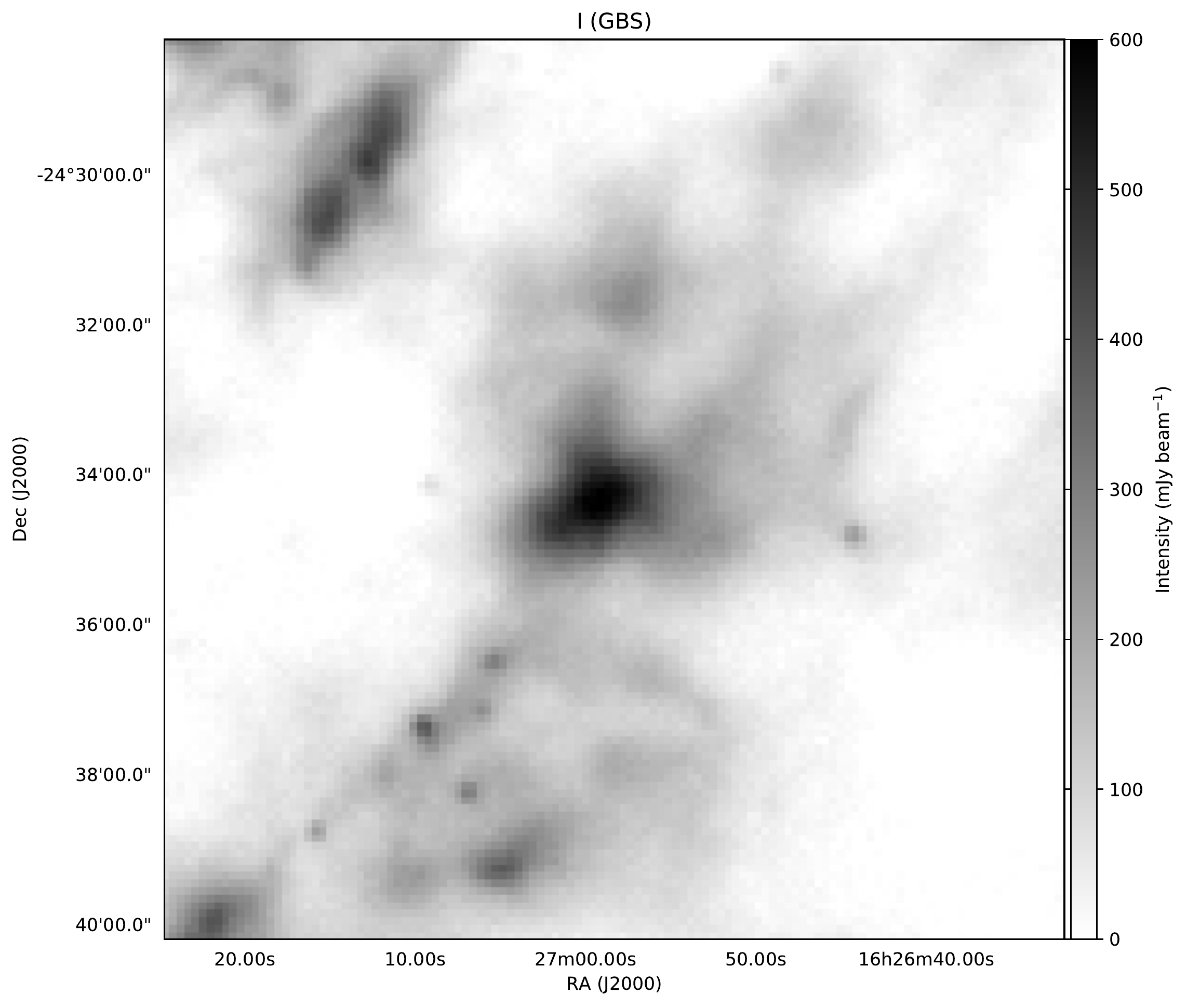}{0.5\textwidth}{(a)}
       \fig{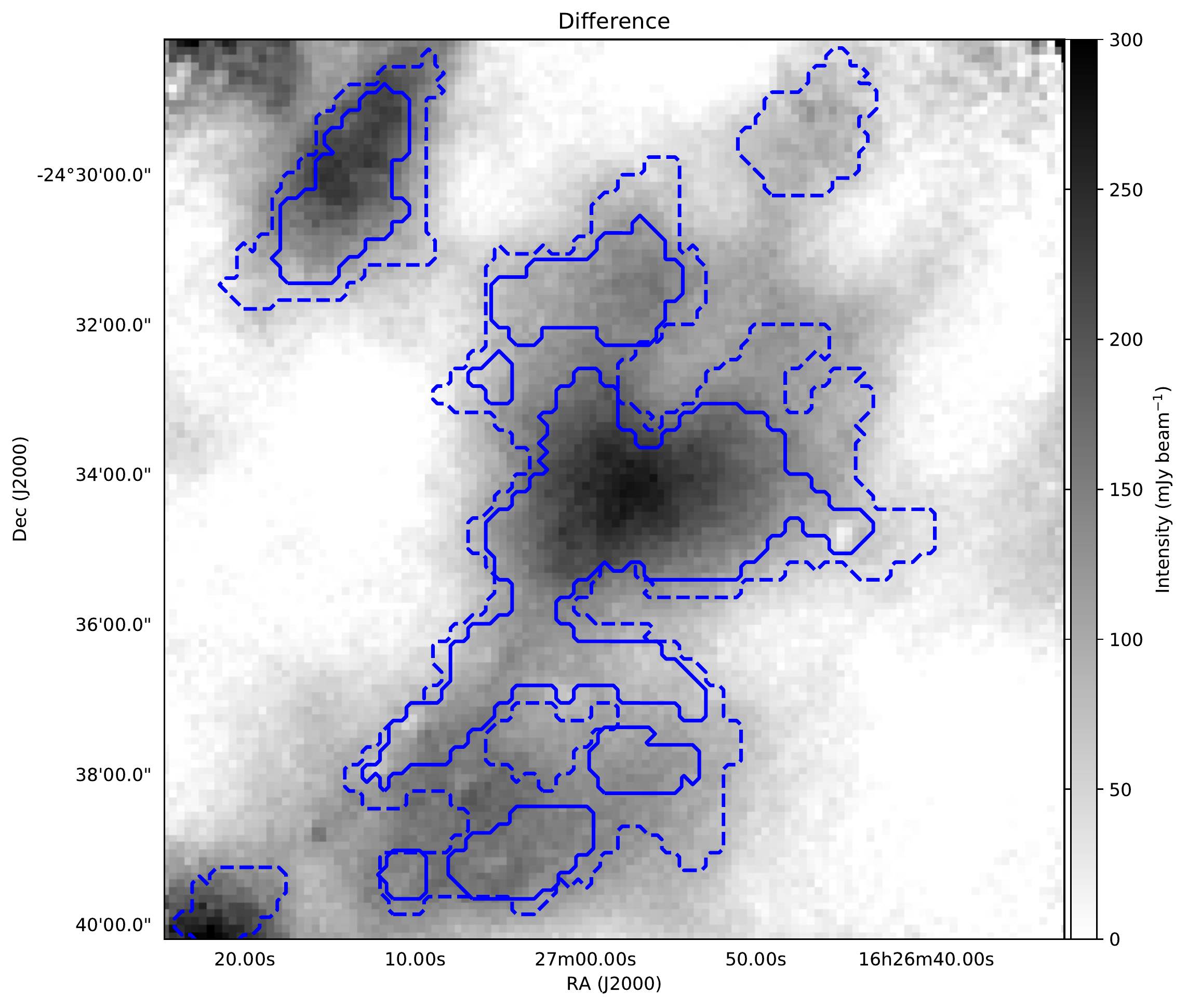}{0.5\textwidth}{(b)}}
\caption{(a) SCUBA-2 850 $\mu$m $I$ map obtained from the GBS project. (b) Difference map obtained by subtracting the POL-2 $I$ map from the SCUBA-2 $I$ map. The intensity is shown in grey scale. The masks are the same as those in Figure \ref{fig:figiqu}  \label{fig:figdiff}}
\end{figure*}

The polarized emission of Oph-C was observed at 850 $\mu$m with SCUBA-2 \citep{2013MNRAS.430.2513H} along with POL-2 \citep[][, Bastien et al. in prep.]{2016SPIE.9914E..03F} between 2016 May 22 and September 10. The region was observed 20 times, among which 19 datasets had an average integration time of 42 minutes and 1 bad dataset was excluded. The observations were made with the POL-2 DAISY mode, which produces a map with high signal-to-noise ratio at the central region of 3$\arcmin$ radius and with increasing noise to the edge. The effective beam size of JCMT is 14.1$\arcsec$ ($\sim$9 mpc at 137 pc) at 850 $\mu$m.

The data were reduced using the SMURF \citep{2013ascl.soft10007J} package in Starlink \citep{2014ASPC..485..391C}. Firstly, the \emph{calcqu} command is used to convert the raw bolometer timestreams into separate Stokes $I$, $Q$, and $U$ timestreams. Then, the \emph{makemap} routine in the \emph{pol2map} script creates individual $I$ maps from the $I$ timestreams of each observation, and coadds them to produce an initial reference $I$ map. Secondly, the \emph{pol2map} is re-run with the initial $I$ map to generate an ASTMASK, which is used to define the signal-to-noise-based background regions that are set to zero until the last iteration, and a PCAMASK, which defines the source regions that are excluded when creating the background models within \emph{makemap}. With the ASTMASK and the PCAMASK, \emph{pol2map} is again re-run to reduce the previously created $I$ timestreams of each observation, creating improved $I$ maps, These individual improved $I$ maps are then coadded to produce a final improved $I$ map. Finally, with the same masks, \emph{pol2map} creates the $Q$ and $U$ maps, along with their variance maps, and the debiased polarization catalogue, from the $Q$ and $U$ timestreams. The final improved $I$ map is used for instrumental polarization correction. The final $I$, $Q$, $U$ maps and the polarization catalogue are gridded to 7$\arcsec$ pixels for a Nyquist sampling. 

The absolute calibration is performed by applying a flux conversion factor (FCF) of 725 Jy beam$^{-1}$ pW$^{-1}$ to the output $I$, $Q$, and $U$ maps, converting the units of these maps from pW to Jy beam$^{-1}$. Due to the additional losses from POL-2, this FCF is 1.35 times larger than the standard SCUBA-2 FCF of 537 Jy beam$^{-1}$ pW$^{-1}$ \citep{2013MNRAS.430.2534D}. The uncertainty on the flux calibration is 5\% \citep{2013MNRAS.430.2534D}. Figure \ref{fig:figiqu} shows the $Q$, $U$, and $I$ maps of our POL-2 data. The rms noises of the background regions in the $Q$ or $U$ maps are $\sim$3.5 mJy beam$^{-1}$. From the corresponding variance maps, the average $Q$ or $U$ variances are $\sim$2 mJy beam$^{-1}$, reaching the target sensitivity value for the BISTRO survey.

Figure \ref{fig:figdiff} shows the total intensity map toward the same region made with SCUBA-2 as part of the Gould Belt Survey (GBS) project \citep{2015MNRAS.450.1094P} and the difference between the SCUBA-2 $I$ map and the POL-2 $I$ map. Because of the difference in the data reduction procedures of the POL-2 data and the SCUBA-2 data and the slower scanning speed of the POL-2 observation, the large-scale structures seen in the SCUBA-2 $I$ map are suppressed in the POL-2 $I$ map. So the BISTRO $I$ map is much fainter than the GBS $I$ map. 

Because of the uncertainties in the $Q$ $\&$ $U$ values and that the polarized intensity and polarized percentage are defined as positive values, the measured polarized intensities are biased toward larger values \citep{2006PASP..118.1340V}. The debiased polarized intensity and its corresponding uncertainty are calculated as:
\begin{equation}
PI = \sqrt{Q^2 + U^2 - 0.5(\delta Q^2 + \delta U^2)},
\end{equation} 
and
\begin{equation}
\delta PI = \sqrt{\frac{(Q^2 \delta Q^2 +  U^2 \delta U^2)}{(Q^2 + U^2)}},
\end{equation} 
where $PI$ is the polarized intensity, $\delta Q$ the uncertainty of $Q$, and $\delta U$ is the uncertainty of $U$. The debiased polarization percentage, $P$ and its uncertainty $\delta P$ are therefore derived by:
\begin{equation}
P = \frac{PI}{I},
\end{equation}
and
\begin{equation}
\delta P = \sqrt{(\frac{\delta PI^2}{I^2} + \frac{\delta I^2 (Q^2 + U^2)}{I^4})},
\end{equation}
where $\delta I$ is the uncertainty of the total intensity. 

Finally, the polarization position angle $\theta$ and its uncertainty $\delta \theta$ \citep{1993A&A...274..968N} are estimated to be:
\begin{equation}
\theta = \frac{1}{2} \tan ^{-1} (\frac{U}{Q}),
\end{equation}
and 
\begin{equation}
\delta \theta = \frac{1}{2} \sqrt{\frac{(Q^2 \delta U^2 +  U^2 \delta Q^2)}{(Q^2 + U^2)^2}},
\end{equation}

\section{Results}

%%The magnetic field in the Oph-C region
\subsection{The magnetic field morphology in the Oph-C region}
\begin{figure*}[htbp]
\centering
\includegraphics[scale=.65]{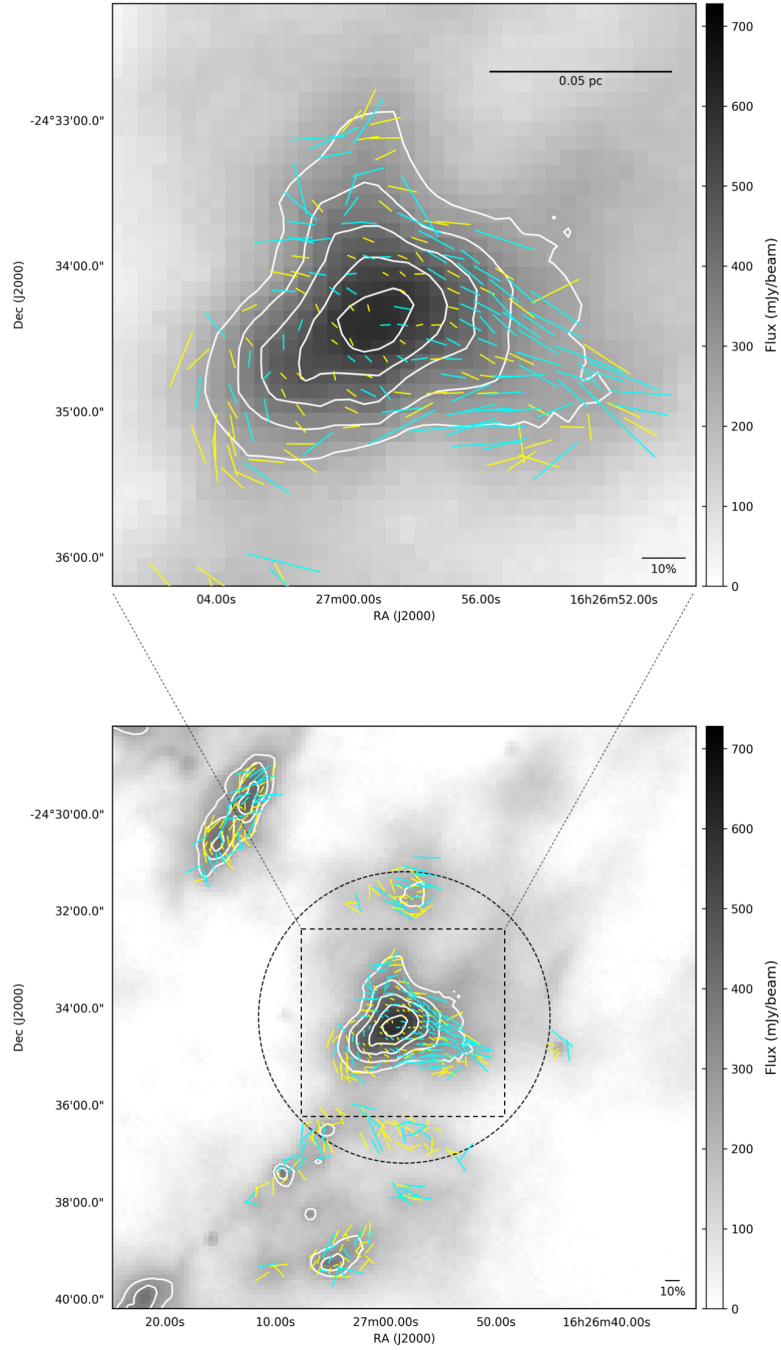}
\caption{Magnetic field orientation maps. The total intensity of the 850 $\mu$m continuum from the GBS project is shown in grey scale. The total intensity is also shown in contour levels, starting from 250 mJy beam$^{-1}$ and continuing at stpdf of 80 mJy beam$^{-1}$. Vectors are from the POL-2 data with $\delta P<5$\%. The yellow and cyan vectors correspond to data with $P/\delta P>2$ and $P/\delta P>3$, respectively. A reference 10\% vector is shown in the lower right. A black dashed circle shows the central region of 3$\arcmin$ radius. \label{fig:figmap}}
\end{figure*}

Assuming that the shortest axis of dust grains is perfectly aligned with the magnetic field, we can derive the orientation of the magnetic field projected on the plane of the sky by rotating the observed polarization vectors by 90$\degr$. Figure \ref{fig:figmap} shows the B-field segments of our POL-2 observations. These segments have lengths proportional to the polarization degrees and orientations along the local B-field. Note that our POL-2 segments are Nyquist sampled with a pixel size of $7''$. With these criteria, the vectors in the Oph-C region are well separated from those in other dense regions of the Ophiuchus cloud. The magnetic field orientations do not appear random, and have a predominant northeast-southwest orientation.

\begin{figure*}[htbp]
\centering
\includegraphics[scale=.65]{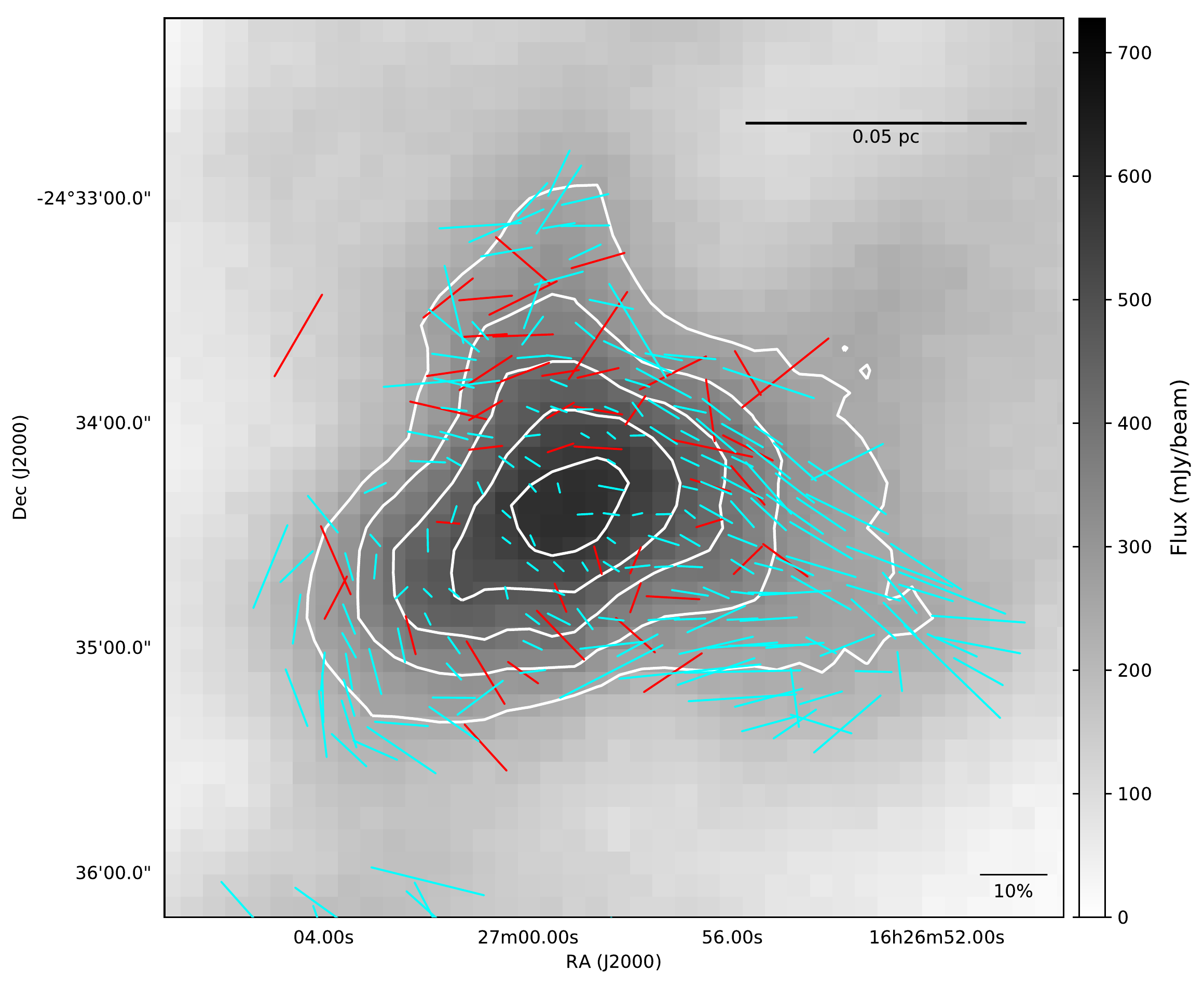}
\caption{Magnetic field orientation maps. The grey scales and contours are the same as those in Figure \ref{fig:figmap}. The cyan and red vectors denote POL-2 data and SCUPOL data where $P/\delta P>2$ and $\delta P<4$\%, respectively. A reference 10\% vector is shown in the lower right. \label{fig:figcom}}
\end{figure*}

Figure \ref{fig:figcom} compares the magnetic field orientations from our POL-2 data with previous observations with the older JCMT polarimeter \citep[SCUPOL,][]{2009ApJS..182..143M}. We use criteria of $P/\delta P>2$ and $\delta P<4$\% for both the POL-2 data and the SCUPOL data. Compared to the previous SCUPOL observations, our POL-2 observations show significant improvements by detecting dust polarization over a much larger area and toward the center of the core. 
%mean 72$\degr$

\begin{figure}[htbp]
\centering
\includegraphics[scale=.45]{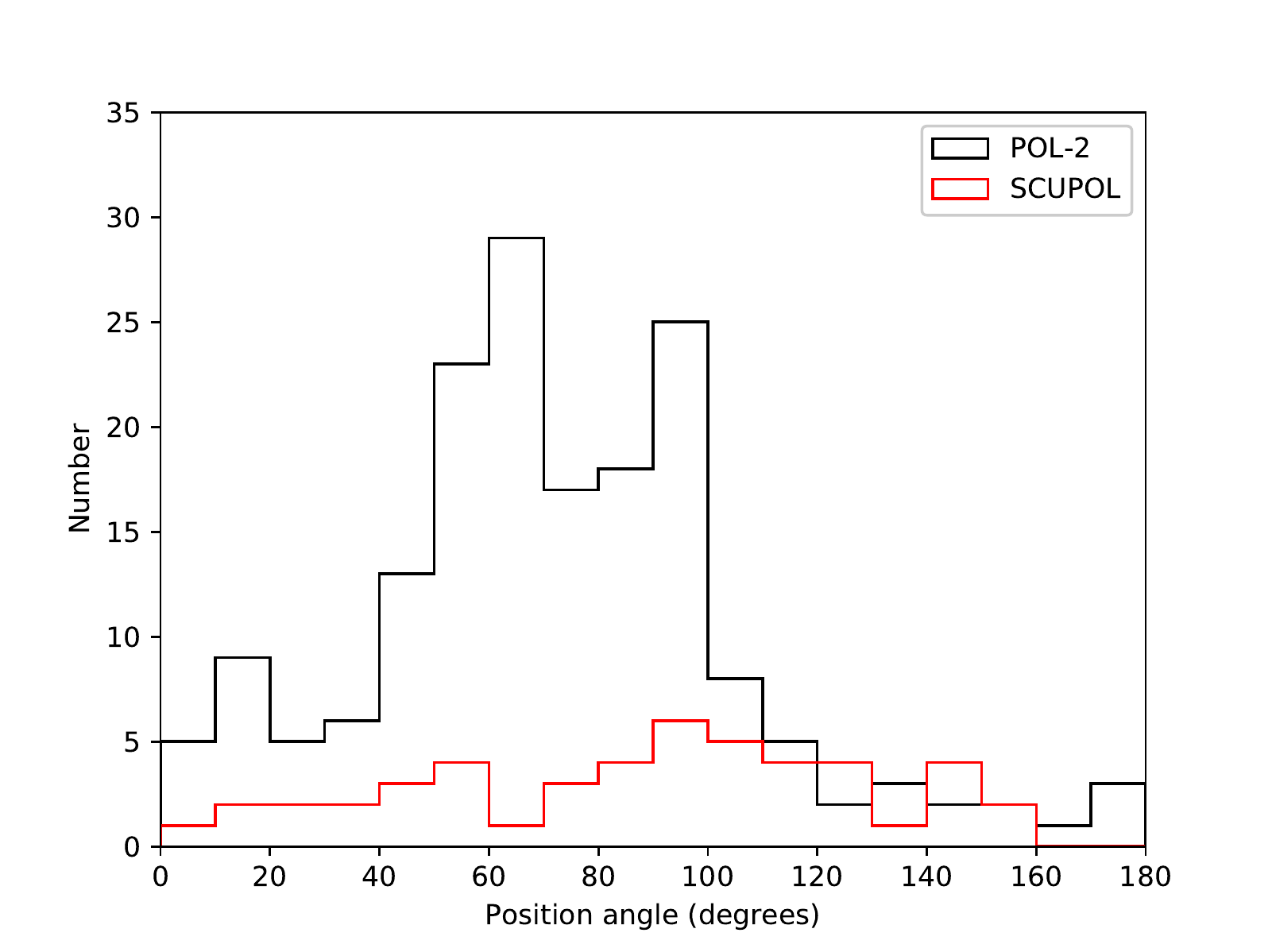}
\caption{Position angle (after 90$\degr$ rotation) histogram for $B$-field vectors with $P/\delta P>2$ and $\delta P<4$\%. The bin size is 10$\degr$. The POL-2 vectors are shown in black. The SCUPOL vectors are shown in red. Angles are measured east of north. \label{fig:figang}}
\end{figure}

In Figure \ref{fig:figang}, histograms of the position angles of the B-field segments from the POL-2 data and the SCUPOL data are shown. The POL-2 histogram has a broad peak between $\sim$40$\degr$ to $\sim$100$\degr$. The standard deviation of the position angles of these POL-2 vectors is $\sim$33$\degr$. The SCUPOL vectors are randomly distributed, which is inconsistent with the POL-2 vectors. 

\begin{figure}[htbp]
\centering
\includegraphics[scale=.45]{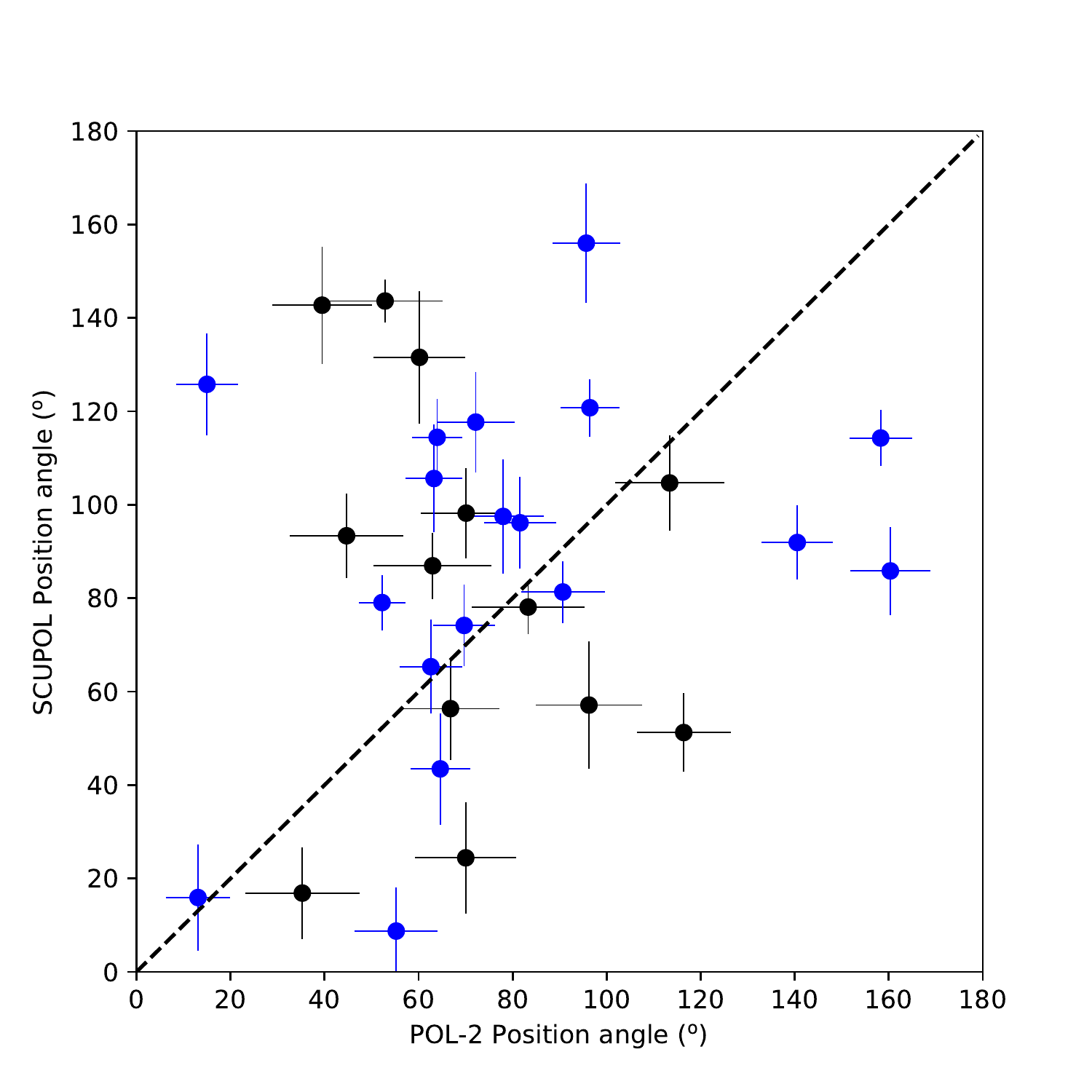}
\caption{Comparision of position angles (after 90$\degr$ rotation) for overlaping SCUPOL and POL-2 vectors with $P/\delta P>2$ and $\delta P<4$\%. Data points correspond to POL-2 data with $P/\delta P>3$ are marked in blue.  Angles are measured east of north. \label{fig:figangang}}
\end{figure}

To further compare the SCUPOL data and the POL-2 data, we resampled the POL-2 data to the same pixel size as that of the SCUPOL data (10$\arcsec$) and aligned the World Co-ordinate System of the two data sets. We found 31 pairs of spatially overlapping vectors between the two data sets with vector selection criteria of $P/\delta P>2$ and $\delta P<4$\%. Figure \ref{fig:figangang} shows the comparision of position angles (after 90$\degr$ rotation) for these overlapping vectors. Large angular differences in the position angles of overlapping vectors can be seen in this figure. The average angular difference of overlapping vectors is estimated to be $\sim$39$\degr$. We further computed the Kolmogorov-Smirnov (K-S) statistic on the POL-2 and SCUPOL position angles and found a probability of 0.06, which suggests the inconsistency in position angles between the two samples. Such a difference, along with the aforementioned inconsistency in histograms of the position angles, can be explained by the lower signal-to-noise ratio of the SCUPOL data.

%%The magnetic field strength
\subsection{The strength of the magnetic field} \label{sec:strength}
\citet{1951PhRv...81..890D} and \citet{1953ApJ...118..113C} proposed that the strength of the B-field could be estimated by interpreting the observed deviation of polarization angles from the mean polarization angle orientation as being due to Alfv\'{e}n waves induced by turbulent perturbations. This interpretation implies that ${\delta}B/B_0\sim\sigma_v/V_{\rm A},$ where ${\delta}B$ is the magnitude of a turbulent component of the B-field, $B_0$ is the strength of the large-scale B-field, $\sigma_v$ is the one-dimensional nonthermal velocity dispersion, and $V_{\rm A}=B_0/\sqrt{4\pi\rho}$ is the Alfv\'{e}n speed for a gas with a mass density of $\rho$ \citep[also see][]{2009ApJ...696..567H}. Such a method (the Davis-Chandrasekhar-Fermi method, DCF method hearafter), in its modified form, has been widely used in estimating the plane-of-sky magnetic field strength $B_{\rm pos}$ from a polarization map by implicitly assuming that ${\delta}B/B_{\rm pos}\sim\sigma_\theta$, where $\sigma_\theta$ is the measured dispersion of polarization angles about a mean or modeled B-field. 

Recently, progress has been made toward more accurately quantifying ${\delta}B/B_{\rm pos}$ from a statistical analysis of polarization angles. In this context, there are different methods based on the ``structure function'' (SF) of polarization angles \citep{2009ApJ...696..567H} or the ``auto-correlation function" (ACF) of polarization angles \citep{2009ApJ...706.1504H}. Yet another approach is to measure the polarization dispersion with a method analogous to ``unsharp masking'' (UM) \citep{2017ApJ...846..122P}. Here we use these methods to estimate $B_{\rm pos}$ in Oph C and compare the results. For the analyses, we use our vector selection criteria of $P/\delta P>3$ and $\delta P<5$\%.

In the original version of DCF's field model \citep{1951PhRv...81..890D, 1953ApJ...118..113C}, the effects of signal integration along line-of-sight and across the beam (hereafter beam-integration effect) were not taken into account. Results from theoretical and numerical works have shown that the beam-integration effect can cause the angular dispersion in polarization maps to be underestimated, therefore overestimating the magnetic field strength \citep{2001ApJ...561..800H, 2001ApJ...546..980O, 2001ApJ...559.1005P, 2008ApJ...679..537F, 2009ApJ...706.1504H, 2016ApJ...821...21C}. To account for this effect, we take a conventional correction factor (we use $Q_c$ to represent this factor throughout this paper) value of 0.5 \citep{2001ApJ...546..980O} to correct the measured angular dispersions and the corresponding magnetic field strength in the SF and UM analyses. The correction parameter $Q_c$ is further discussed in Section \ref{methods}.

Assuming the optically thin dust emission, a dust-to-gas ratio $\Lambda$ of 1:100 \citep{1991ApJ...381..250B}, and an opacity index $\beta$ of 2 \citep{1983QJRAS..24..267H}, we calculate gas column density $N(H_2)$ following:
\begin{equation}
N(H_2) = \frac{I_{\nu}}{\mu m_{\mathrm{H}} \kappa_{\nu} B_{\nu} (T)},
\end{equation}
where $I_{\nu}$ is the continuum intensity at frequency $\nu$, $\mu = 2.86$ is the mean molecular weight \citep{2013MNRAS.432.1424K, 2015MNRAS.450.1094P}, $m_{\mathrm{H}}$ is the atomic mass of hydrogen,  $\kappa_{\nu} = 0.1 (\nu / 1 \mathrm{THz})^{\beta}$ is the dust opacity \citep{1983QJRAS..24..267H} in cm$^2$ g$^{-1}$, and $B_{\nu} (T)$ is the Planck function at temperature $T$. In our analyses, we adopted a dust temperature of 10 $\pm$ 3 K \citep{2007MNRAS.379.1390S}. The uncertainty on the estimation of column density mainly comes from the uncertainty of $\kappa_{\nu}$ \citep{1995P&SS...43.1333H}. Conservatively, we adopt a fractional uncertainty of 50\% \citep{2014A&A...562A.138R, 2017ApJ...846..122P} for $\kappa_{\nu}$. In our calculations, we ignore the uncertainties on $\Lambda$ and $\mu$. The column density was estimated over the area with Stokes $I>250$ mJy beam$^{-1}$ (see Figure \ref{fig:figmap}). The measured area $A$ is 14544 arcsec$^2$ (0.0053 pc$^2$). Taking into account the uncertainties on $\kappa_{\nu}$, temperature, and flux calibration, the fractional uncertainty on the estimated column density is 59\%. Therefore, the mean column density in the concerned region is estimated to be (1.05 $\pm$ 0.62) $\times$ 10$^{23}$ cm$^{-2}$. Since the Oph-C core is highly `centrally condensed' \citep{1998A&A...336..150M}, we adopt a spherical geometry and a core volume ($V$) of 4/3($A^3$/$\pi$)$^{1/2}$. Again, we ignore the uncertainty on the geometry assumption. The average volume density $n_{\mathrm{H_2}}$ is estimated to be (6.4 $\pm$ 3.7) $\times$ 10$^{5}$ cm$^{-3}$. The total mass in our measured volume, $M = \mu m_\mathrm{H} N (\mathrm{H_2}) A$, is 12 $\pm$ 7 solar masses.

To calculate the plane-of-sky magnetic field strength, we need information about the velocity dispersion of the gas. Assuming isotropic velocity perturbations, we adopt the line-of-sight velocity dispersion estimated by \citet{2007A&A...472..519A}. In their work, they carried out N$_2$H$^+$ (1--0) observations toward the Ophiuchus main cloud with a 26$\arcsec$ beam using the IRAM 30m telescope, and found that the average line-of-sight nonthermal velocity dispersion, $\sigma_v$, of the dense structures in Oph-C is 0.13 $\pm$ 0.02 km s$^{-1}$. Their N$_2$H$^+$ data are appropriate to trace the velocity dispersion in Oph-C because of many reasons. At 10 K, the critical density of N$_2$H$^+$ (1--0) is 6.1 $\times$ 10$^4$ cm$^{-3}$ \citep{2015PASP..127..299S}, which is sufficient to probe the dense materials in Oph-C. Also the masses of dense structures in Oph-C traced by the N$_2$H$^+$ (1--0) data and the SCUBA-2 850 $\mu$m continuum data are in good agreement \citep{2015MNRAS.450.1094P} and the N$_2$H$^+$ (1--0) in Oph-C is optically thin \citep{2007A&A...472..519A}, indicating the N$_2$H$^+$ data and our SCUBA-2/POL-2 data generally trace the same material. Although the beam size of the N$_2$H$^+$ observation is nearly twice the beam size of our POL-2 observation, the spatial resolution of the N$_2$H$^+$ observation, 26$\arcsec$, is still sufficient to resolve the Oph-C core that has a diameter of $\sim$2$\arcmin$. Thus, it could be concluded that the average line-of-sight nonthermal velocity dispersion of the dense structures in Oph-C traced by N$_2$H$^+$ (1--0) is well suited to represent the average gas motions in our concerned region. 

\begin{table*}[]
\centering
\begin{tabular}{c|l|c|c|c}\hline
Parameter &Description&SF&ACF &UM\\\hline
$\Delta \theta$ (degrees)&Angular dispersion&22 $\pm$ 1 &21 $\pm$ 8	&11 $\pm$ 1\\\hline
$\langle  \delta B^2 \rangle $/$ \langle B_0^2 \rangle$	&Turbulent-to-ordered magnetic field energy ratio&0.15 $\pm$ 0.01	&0.13 $\pm$ 0.10	&0.035 $\pm$ 0.004	\\\hline
$B_{\rm pos}$ ($\mu$G)&Plane-of-sky magnetic field strength& 206 $\pm$ 68	& 223 $\pm$ 113	& 426 $\pm$ 141\\\hline

\end{tabular}
\caption{Parameters derived from different modified DCF methods without correction for beam-integration. \label{tab:para_int}}
\end{table*}

\begin{table*}[]
\centering
\begin{tabular}{c|l|c|c|c}\hline
Parameter &Description&SF&ACF &UM\\\hline
$\Delta \theta$ (degrees)&Angular dispersion&45 $\pm$ 14	&34 $\pm$ 13	&21 $\pm$ 7\\\hline
$\langle  \delta B^2 \rangle $/$ \langle B_0^2 \rangle$	&Turbulent-to-ordered magnetic field energy ratio&0.61 $\pm$ 0.37	&0.35 $\pm$ 0.27	& 0.14 $\pm$ 0.09\\\hline
$B_{\rm pos}$ ($\mu$G)&Plane-of-sky magnetic field strength& 103 $\pm$ 46	& 136 $\pm$ 69	& 213 $\pm$ 115\\\hline
 $\lambda$ &Observed magnetic stability critical parameter&7.8 $\pm$ 5.7		&5.9 $\pm$ 4.6	&3.8 $\pm$ 3.0\\\hline
 $\lambda_c$ &Corrected magnetic stability critical parameter&2.6 $\pm$ 1.9		&1.9 $\pm$ 1.5	&1.3 $\pm$ 1.0\\\hline
$E_{\mathrm{B}}$ ($10^{35}$ J)&Total magnetic energy  & 5.4 $\pm$ 4.8 & 9.5 $\pm$ 9.7 & 23.2 $\pm$ 25.0 \\\hline
\end{tabular}
\caption{Parameters derived from different modified DCF methods with correction for beam-integration. \label{tab:para}}
\end{table*}

%Structure function analysis
\subsubsection{Structure function analysis}
In the SF method \citep{2009ApJ...696..567H}, the magnetic field is assumed to consist of a large-scale structured field, $B_0$, and a turbulent component,  ${\delta}B$. The structure function infers the behavior of position angle dispersion as a function of vector separation $l$. At some scale larger than the turbulent scale $\delta$,  ${\delta}B$ should reach its maximum value. At scales smaller than a scale $d$, the higher-order terms of the Taylor expansion of $B_0$ can be cancelled out. When $\delta < l \ll d$, the angular dispersion function follows the form:
\begin{equation}
\langle \Delta \Phi ^2 (l)\rangle_{\mathrm{tot}} \simeq b^2 + m^2l^2+\sigma_M^2(l).
\end{equation}
In this equation, $\langle \Delta \Phi ^2 (l)\rangle_{\mathrm{tot}}$, the square of the total measured dispersion function, consists of $b^2$, a constant turbulent contribution, $m^2l^2$, the contribution from the large-scale structured field, and $\sigma_M^2(l)$, the contribution of the measurement uncertainty. The ratio of the turbulent component and the large-scale component of the magnetic field is given by:
\begin{equation}
\frac{\langle  {\delta}B^2 \rangle^{1/2}}{B_0} = \frac{b}{\sqrt{2-b^2}}.
\end{equation}
And $B_0$ is estimated according to: 
\begin{equation}
B_0 \simeq \sqrt{(2-b^2)4\pi \mu m_{\mathrm{H}} n_{\mathrm{H_2}}} \frac{\sigma_v}{b}.
\end{equation}
Then the estimated plane-of-sky magnetic field strength is corrected by $Q_c$:
\begin{equation}
B_{\mathrm{pos}} = Q_cB_0.
\end{equation}

\begin{figure}[htbp]
\centering
\includegraphics[scale=.45]{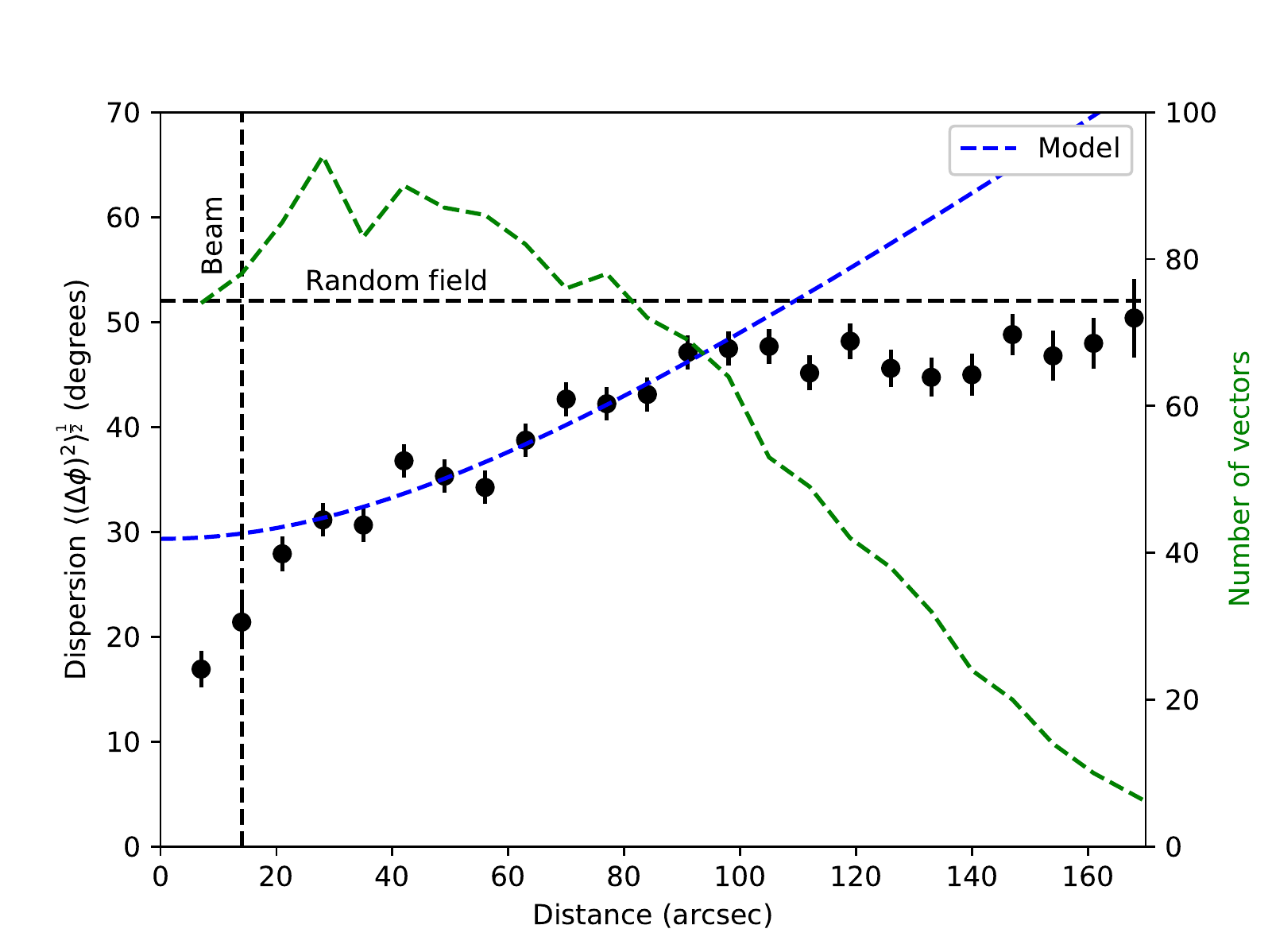}
\caption{Left-hand axis: Angular dispersion function for Oph-C. The angle dispersion segments are shown in black solid circles with error bars. The measurement uncertainties were removed. The best fit is shown by the blue dashed line. Right-hand axis: the number of independent vectors (green dashed line) included in each data bin. The bin size is 7$\arcsec$. \label{fig:figstru}}
\end{figure}

Figure \ref{fig:figstru} shows the angular dispersion corrected by uncertainty ($\langle \Delta \Phi ^2 (l)\rangle_{\mathrm{tot}} - \sigma_{M}^2(l)$) as a function of distance measured from the polarization map. Following \citet{2009ApJ...696..567H}, the data are divided into separate distance bins with separations corresponding to the pixel size. At scales of 0-25$\arcsec$, the angular dispersion function increases steeply with the segment distance, most possibly due to the contribution of the turbulent field. At scales larger than 25$\arcsec$, the function continues increasing with a shallower slope, which we may attribute to the large-scale ordered magnetic field structure, and reaches its maximum at $\sim$100$\arcsec$. The maximum of the angular dispersion function is lower than the value expected for a random field \citep[52$\degr$,][]{2010ApJ...716..893P}. The angular dispersion function presents wave-like ``jitter'' features at $l >$25$\arcsec$. \citet{2016A&A...596A..93S} have attributed the jitter features to the sparse sampling of the vectors in the observed region, which means the independent vectors involved in each distance bin are not enough to achieve statistical significance. We performed simple Monte Carlo simulations (see Appendix \ref{modeling}) and found that the uncertainty from sparse sampling is $\sim$1.5$\degr$ in the structure function for models with SFs similar to that of our data in amounts of large-scale spatial correlation and random angular dispersions. We fit the structure function over $25\arcsec<l<100\arcsec$. During the fitting, both the uncertainties from the sparse sampling and from simply propagating the measurement uncertainties of the observed position angles have been taken into account. The reduced chi-squared ($\chi_{red}^2$) of the fitting is 1.2. The calculated values of parameters are given in Table \ref{tab:para_int} (without correction for the beam-integration effect) and Table \ref{tab:para} (with correction for the beam-integration effect).

%Correlation function analysis
\subsubsection{Auto-correlation function analysis}

The ACF method \citep{2009ApJ...706.1504H} expands the SF method by including the effect of signal integration along the line of sight and within the beam in the analysis. \citet{2009ApJ...706.1504H} write the angular dispersion function in the form:
\begin{equation}
1 - \langle \cos \lbrack \Delta \Phi (l)\rbrack \rangle \simeq \frac{1}{N} \frac{\langle {\delta}B^2 \rangle}{\langle B_0^2 \rangle} \times \lbrack 1 - e^{-l^2/2(\delta^2+2W^2)}\rbrack + a_2\arcmin l^2,
\end{equation}
where $\Delta \Phi (l)$ is the difference in position angles of two vectors seperated by a distance $l$, $W$ the beam radius (6.0$\arcsec$ for JCMT, i.e., the FWHM beam divided by $\sqrt{8 \ln{2}}$), $a_2\arcmin$ is the slope of the second-order term of the Taylor expansion, and $\delta$ is the turbulent correlation length mentioned before. $N$ is the number of turbulent cells probed by the telescope beam and is given by: 
\begin{equation}
N = \frac{(\delta^2 + 2W^2)\Delta\arcmin}{\sqrt{2\pi}\delta^3},
\end{equation}
where $\Delta\arcmin$ is the effective thickness of the cloud. The ordered magnetic field strength can be derived by: 
\begin{equation}
B_0 \simeq \sqrt{4\pi \mu m_{\mathrm{H}} n_{\mathrm{H_2}}} \sigma_v \left[ \frac{\langle  {\delta}B^2 \rangle}{\langle B_0^2 \rangle} \right]^{-1/2}.
\end{equation}

\begin{figure}[!tbp]
\gridline{\fig{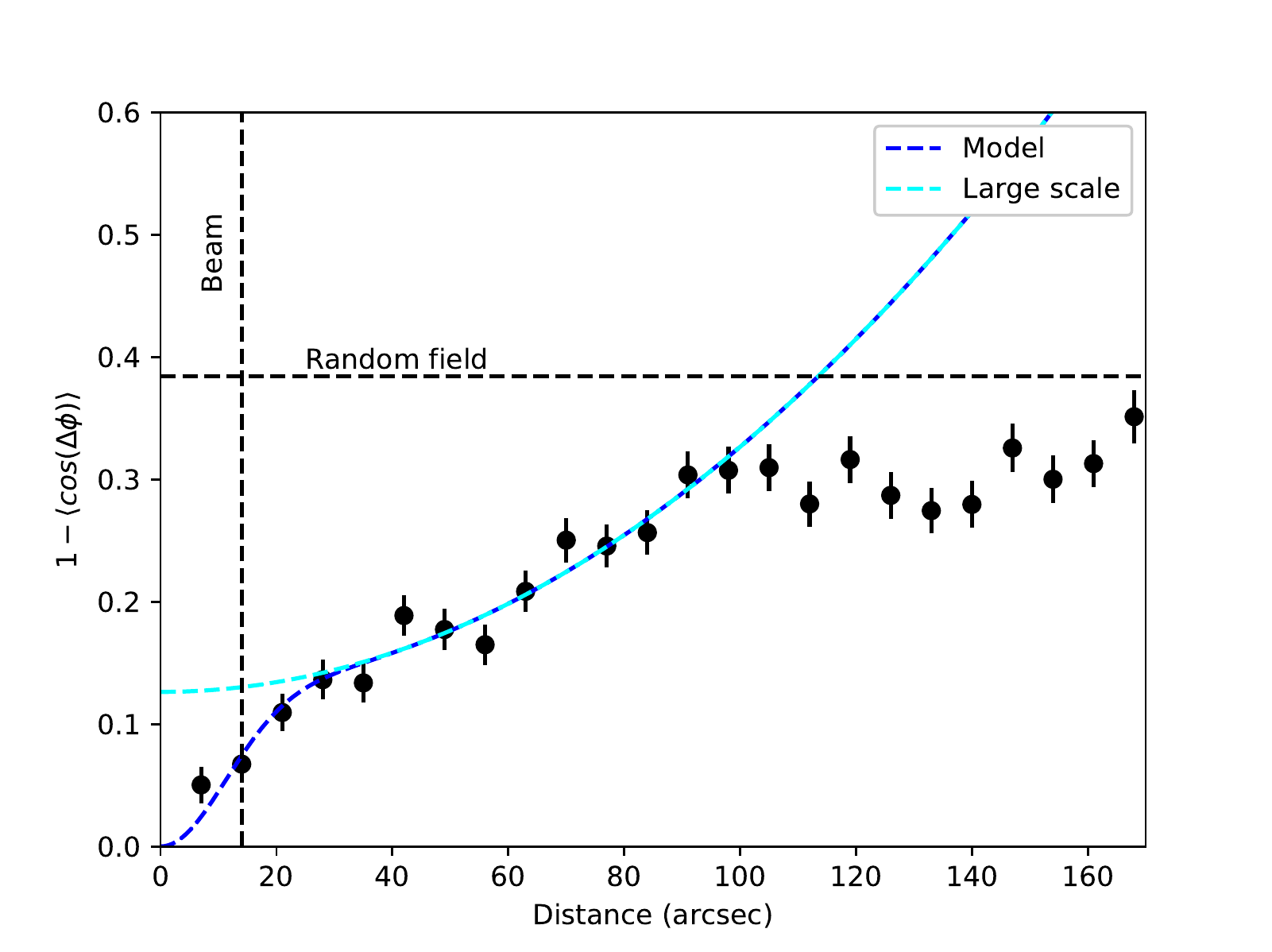}{0.45\textwidth}{(a)}
      }
 \gridline{\fig{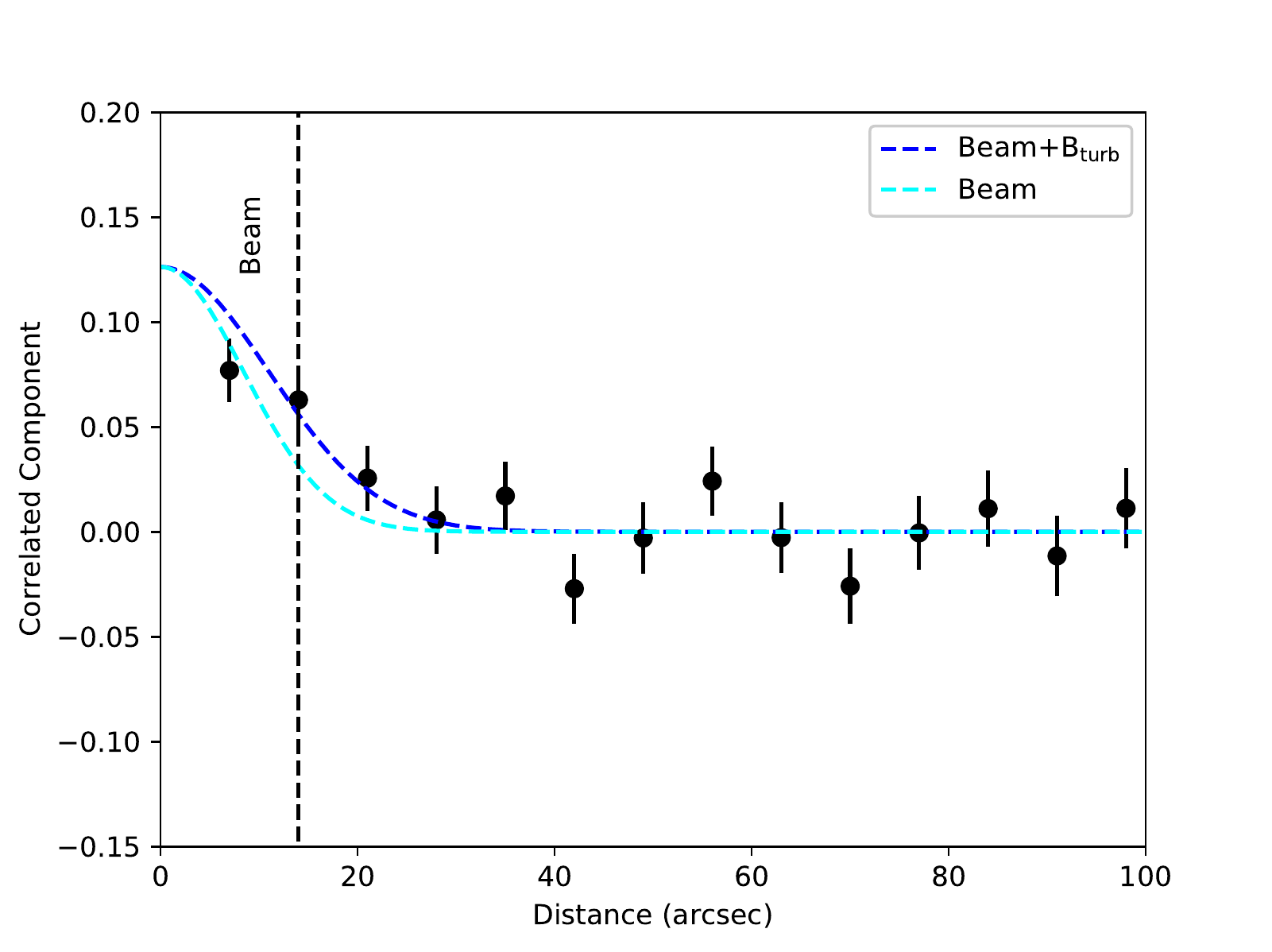}{0.45\textwidth}{(b)}
      }
\caption{(a) Angular dispersion function for Oph-C. The angle dispersion segments are shown in black solid circles with error bars. The bin size is the same with that in Figure \ref{fig:figstru}. A blue dashed line shows the fitted dispersion function. The cyan dashed line shows the large-scale component $(1/N) (\langle {\delta}B^2 \rangle / \langle B_0 \rangle) + a_2\arcmin l^2$ of the best fit. (b) Correlated component of the dispersion function. The correlated component $(1/N) \langle  {\delta}B^2 \rangle / \langle B_0 \rangle)  e^{-l^2/2(\delta^2+2W^2)}$ is shown in blue dashed line. The cyan line shows the correlated component solely due to the beam. \label{fig:figcorre}}
\end{figure}

Figure \ref{fig:figcorre}(a) shows the angular dispersion function of the polarization segments in the Oph-C region. Figure \ref{fig:figcorre}(b) shows the correlated component of the dispersion function. The uncertainty from sparse sampling is $\sim$0.015 in the auto-correlation function (see Appendix \ref{modeling}). We fit the function at $l<100\arcsec$. Again, both the uncertainties from the sparse sampling and from the measurements have been taken into account. In our fitting, $\Delta \arcmin$ is set to 20$\arcsec$, which is roughly the FWHM of the starless sub-core identified by \citet{2015MNRAS.450.1094P}. The $\chi_{red}^2$ of the fitting is 1.1. The turbulent correlation length $\delta$ is found to be 7.0$\arcsec$ $\pm$ 2.7$\arcsec$ (4.3 $\pm$ 1.6 mpc). The number of turbulent cells is derived to be 2.5 $\pm$ 0.5. The calculated values of other parameters are given in Tables \ref{tab:para_int} and \ref{tab:para}.

%Unsharp masking analysis
\subsubsection{Unsharp masking analysis}
In this section, we followed \citet{2004ApJ...600..279C} to derive the plane-of-sky magnetic field strength with the expression:
\begin{equation}\label{eq:eqcf}
B_{\mathrm{pos}} = Q_c \sqrt{4\pi \mu m_{\mathrm{H}} n_{\mathrm{H_2}}}\frac{\sigma_\mathrm{v}}{\sigma_{\mathrm{\theta}}}.
\end{equation} 
The dispersion of the magnetic field angle, $\sigma_{\mathrm{\theta}}$, is measured following the unsharp masking method developed by \citet{2017ApJ...846..122P}. Firstly, a 3 $\times$ 3 pixel boxcar average is applied to the measured angles to show the local mean field orientation. With a 3 $\times$ 3 pixel boxcar, the effect of the curvature of the large-scale ordered field on the smoothing is minimized. Then the deviation in field angle from the mean field orientation is derived by subtracting the smoothed map from the observed magnetic field map. Finally, the standard deviation of the residual angles is measured to represent the angular dispersion of the magnetic field angle. 

\begin{figure*}[!tbp]
\centering
\includegraphics[scale=.6]{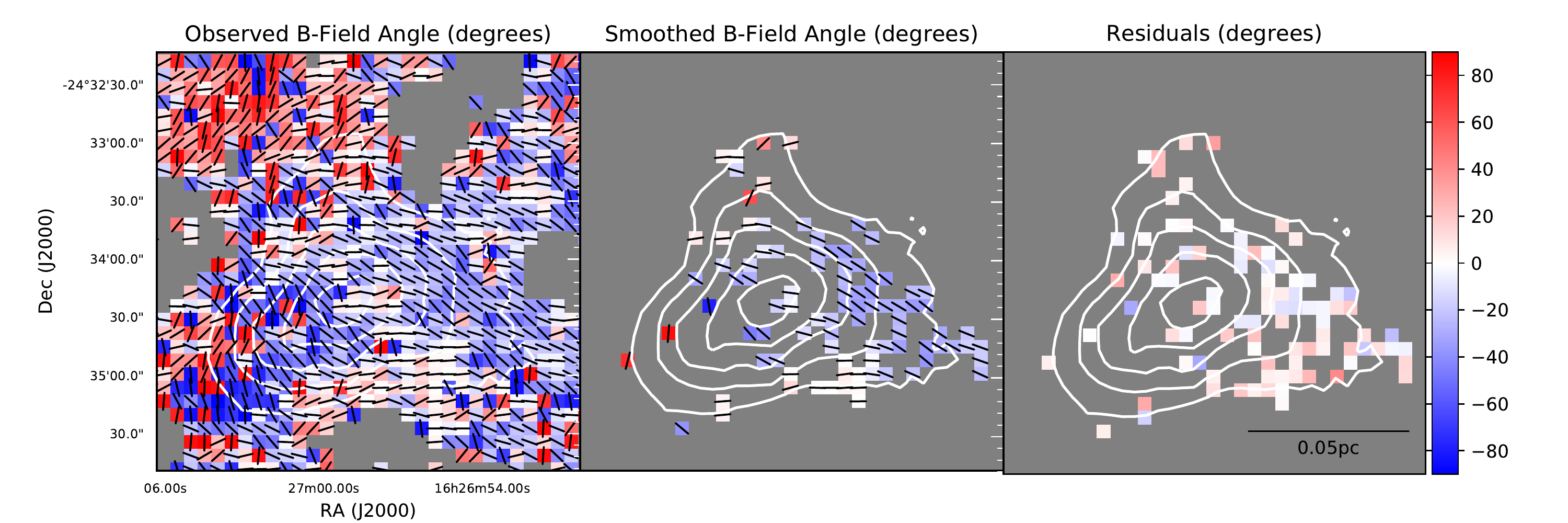}
\caption{Left panel: The observed magnetic field angles $\theta_{\mathrm{obs}}$. Middle panel: The smoothed magnetic field angles $\langle \theta \rangle$. Right panel: The residual angles $\theta_{\mathrm{obs}} - \langle \theta \rangle$. The vectors are of uniform length. The angles are measured south of east so that the color bars of the observed, smoothed, and residual angles are unified. \label{fig:figunsharp}}
\end{figure*}

\begin{figure*}[!tbp]
\centering
\includegraphics[scale=.6]{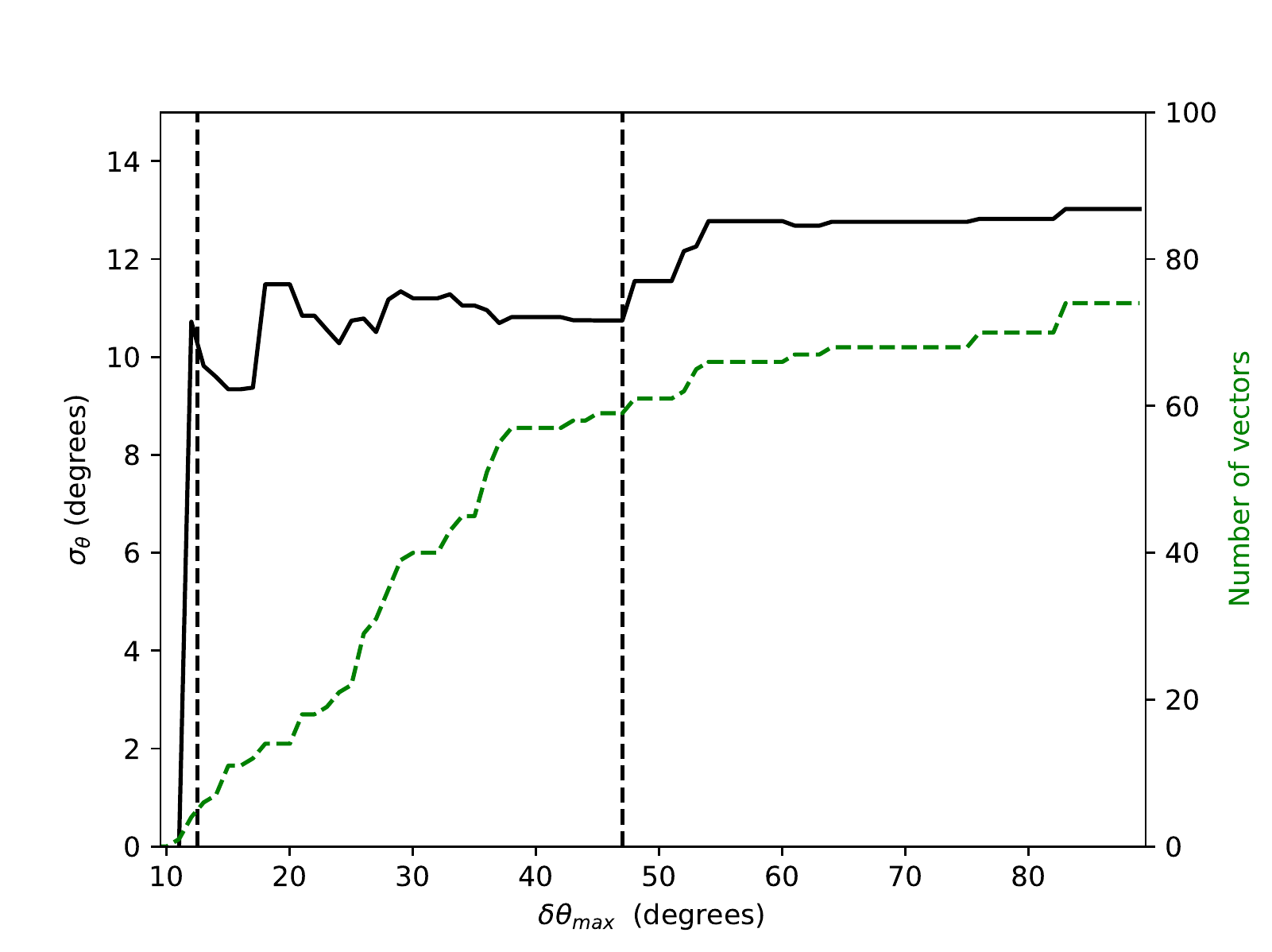}
\caption{Left-hand axis: $\sigma_{\theta}$ as a cumulative function of $\delta \theta_{\mathrm{max}}$ is shown in black line. Right-hand axis: the number of vectors (green dashed line) included in the cumulative function. The average standard deviation is measured over the region between two black dashed lines. \label{fig:figcumu}}

\end{figure*}

We applied the UM method on our data and restricted the analysis to pixels where the maximum angle difference within the boxcar is $< 90 \degr$. Figure \ref{fig:figunsharp} shows the observed position angles $\theta_{\rm obs}$, the position angles $<\theta>$ of a mean B-field derived by smoothing the observed position angles with a $3\times3$ pixel boxcar filter, and the residual values $\theta_{\rm obs}-<\theta>$. We then calculated the standard deviation of magnetic field angles ($\sigma_{\theta}$) as a cumulative function of the maximum permitted angle uncertainty ($\delta \theta_{\mathrm{max}}$) in the 3 $\times$ 3 pixel smoothing box (see Figure \ref{fig:figcumu}). With Monte Carlo simulations, \citet{2017ApJ...846..122P} found that $\sigma_{\theta}$ can well represent the true angular dispersion when $\delta \theta_{\mathrm{max}}$ is small, while $\sigma_{\theta}$ tend to increase with $\delta \theta_{\mathrm{max}}$ when $\delta \theta_{\mathrm{max}}$ is large. In our case, we restrict our analysis to 12$\degr < \delta \theta_{\mathrm{max}} <$ 47$\degr$, where $\sigma_{\theta}$ remains relatively constant within this $\delta \theta_{\mathrm{max}}$ range. The average standard deviation is measured to be 10.7$\degr \pm$0.6$\degr$ (see Figure \ref{fig:figcumu}). This value is introduced in Equation \ref{eq:eqcf} as $\sigma_{\mathrm{\theta}}$. The calculated values of other parameters are given in Tables \ref{tab:para_int} and \ref{tab:para}.

\section{Discussion} 
\subsection{Structure and orientation of the magnetic field}
Oph-C is unique in the Ophiuchus cloud as it is fully a starless core. Investigating the magnetic field structure in starless cores is essential for us to explore the initial conditions of star formation. Previous polarization observations toward cores in the starless phase have shown relatively smooth and uniform magnetic field structures \citep{2000ApJ...537L.135W, 2004ApJ...600..279C, 2009MNRAS.398..394W}. Recently, \citet{2017ApJ...845...32K} presented the first detection of an hourglass-shaped magnetic field in a starless core with NIR polarization observations toward FeSt 1–457 (also known as Pipe-109), suggesting that the magnetic field lines can be distorted by mass condensation in the starless phase. However, the NIR polarization observations cannot trace the densest materials in the core and the hourglass morphology was not found in sub-mm polarization observations toward the same source \citep{2014A&A...569L...1A}. Our observations toward Oph-C, which present the most sensitive sub-mm polarization observation in a low-mass starless core to date, reveal an relatively ordered B-field with a prevailing northeast-southwest orientation (see Figure \ref{fig:figmap}). However, the B-field structure in Oph-C shows no evidences of an hourglass morphology, which is consistent with previous observations that an hourglass morphology is not generally found in other cold dense cores from sub-mm polarization observations at scales $>$ 0.01 pc. This suggests that mass condensation does not significantly distort the local B-field structure at scales $>$ 0.01 pc in the densest materials of dense cores at both the starless phase and prestellar phase.

The role of magnetic field in dense cores may vary with the evolution of the core. As part of the BISTRO survey \citep{2017ApJ...842...66W}, polarization observations towards two protostellar cores with similar masses \citep{1998A&A...336..150M} as that of Oph-C in the Ophiuchus cloud, Oph-A and Oph-B, have been made and are ready to be compared with our data. Our observations of Oph-C show that the overall magnetic field geometry in Oph-C is ordered and the polarization position angles show large angular dispersions. This behaviour is similar with that in Oph-B, which is relatively a quiescent core in Ophiuchus but is more evolved than Oph-C, while the B-field in Oph-A, which is the warmest and the only core with substantially gravitationally bound sub-cores found in Ophiuchus, is mostly well organized and with small angular dispersions \citep{2009ApJ...692..973E, 2015MNRAS.450.1094P, 2018ApJ...859....4K, 2018ApJ...861...65S}. In addition, the angular dispersions in Oph-C (($\sim$11$\degr$ to $\sim$22 $\degr$)) and Oph-B ($\sim$15$\degr$) is larger than that in Oph-A ($\sim$2 to $\sim$6$\degr$). These indicate that the star formation process may possibly reduce angular dispersions in the magnetic field in the late stages of star formation. 

Our observations reveal that there is a prevailing orientation in the B-field in Oph-C centering at $\sim$40$\degr$ to $\sim$100$\degr$ (see Figure \ref{fig:figang}). This orientation agrees with the B-field orientations in Oph-A, where the B-field components center at $\sim$40$\degr$ to $\sim$100$\degr$ \citep{2018ApJ...859....4K}, and Oph-B, where the position angle of B-field peaks at $\sim$50$\degr$ to $\sim$80$\degr$ \citep{2018ApJ...861...65S}. The B-field position angle distribution in Oph-C is consistent with the $\sim$50$\degr$ B-field component in lower-density regions of the Ophiuchus cloud traced by the NIR polarization map of \citet{2015ApJS..220...17K}, and is also aligned with the cloud-scale B-field orientation probed by Planck  \citep[see Figure 3, ][]{2016A&A...586A.138P}. The consistence of B-field orientation from cloud to core scales indicates that the large-scale magnetic field plays a dominant role in the formation of dense cores in the Ophiuchus cloud. 

\subsection{Depolarization effect}
\begin{figure}[!tbp]
\gridline{\fig{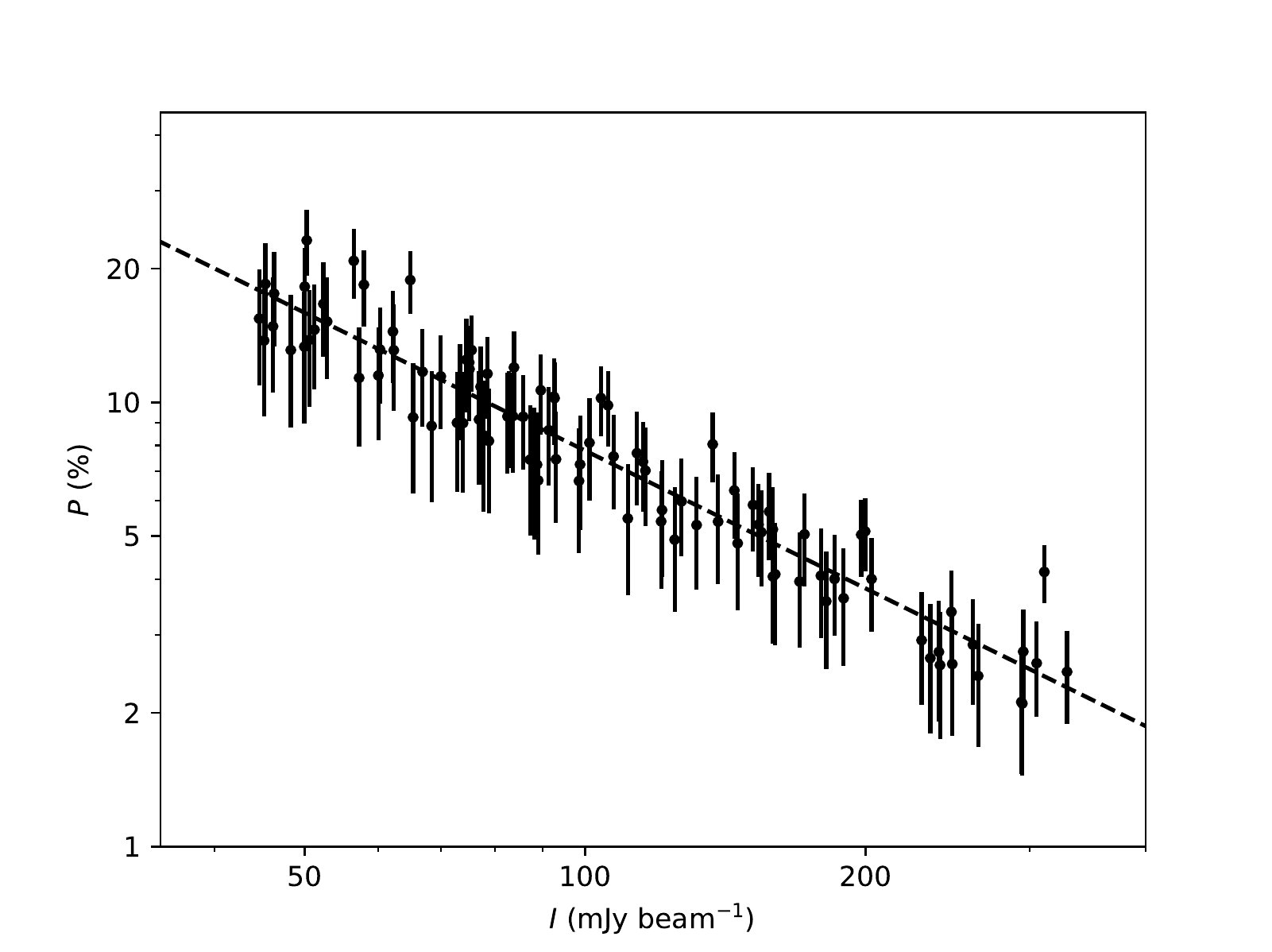}{0.45\textwidth}{(a)}
      }
 \gridline{\fig{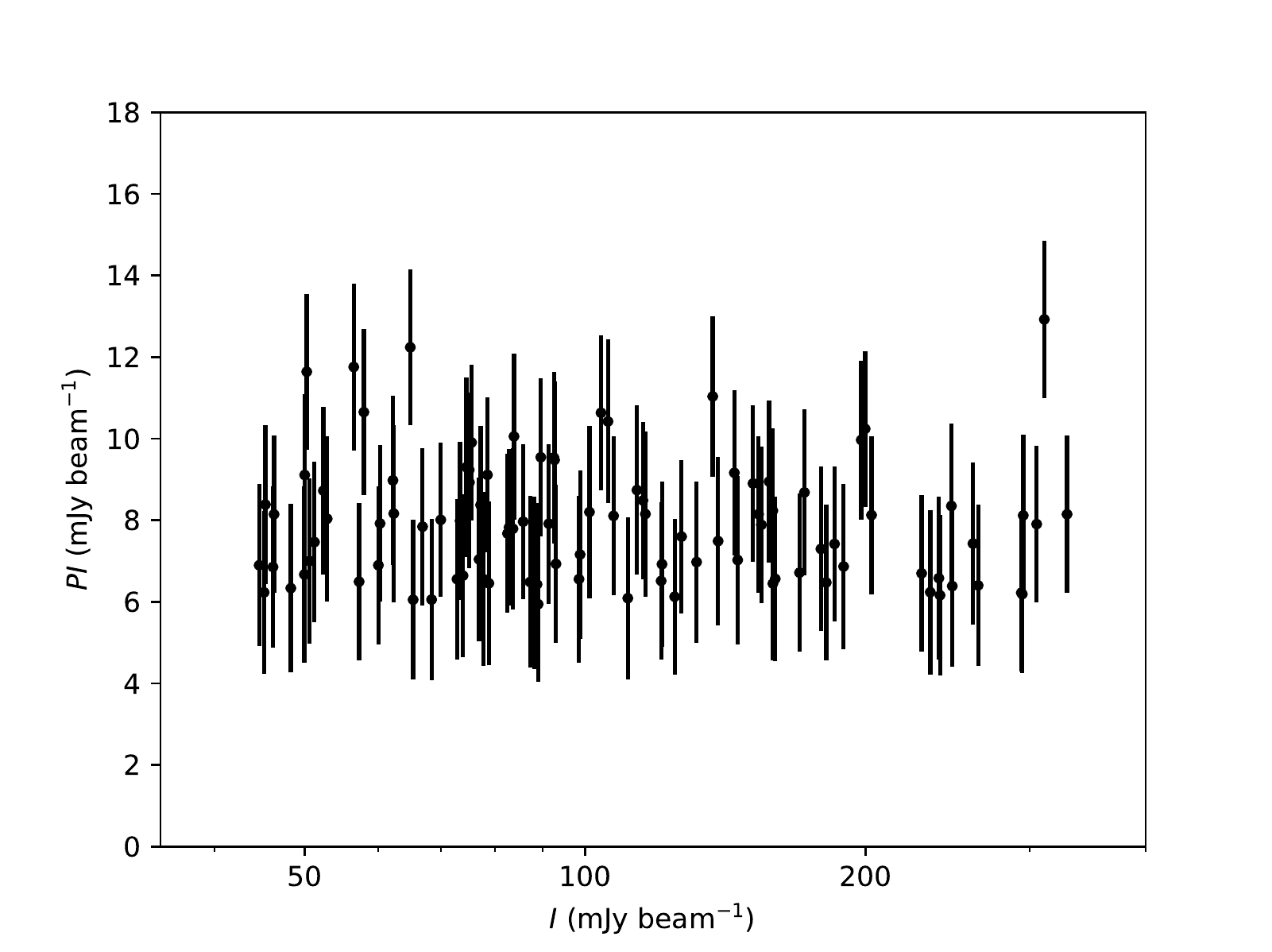}{0.45\textwidth}{(b)}
      }
\caption{(a) Polarization degree vs. total intensity. (b) Polarized intensity vs. total intensity. Data points with $P/\delta P>3$ and $\delta P<5$\% are shown with error bars. The result of the power-law fitting of the $P$-$I$ relation is shown in dashed line. \label{fig:figpi}}
\end{figure}

A clear trend of decreasing polarization percentage with increasing dust emission intensity is seen in Figure~\ref{fig:figmap}. Such an effect is more evident in Figure~\ref{fig:figpi}(a), in which the $P$-$I$ relation suggests depolarization toward high density. Considering that the overall field does not change orientation while threading the core, the depolarization in Oph-C seems unlikely to be a by-product of field tangling of complex small-scale field lines within the JCMT beam. In addition, the subsonic nonthermal gas motions of Oph-C indicate that the polarization percentage has not been significantly affected by the number of turbulent cells along line of sight. The RAT mechanism \citep{2007JQSRT.106..225L}, which suggests inefficient grain alignment toward high density regions, also cannot fully explain the depolarization effect because of the lack of an internal or external radiation field in the Oph-C region. Alternatively, grain characteristics such as size, shape, and composition, which are related to grain alignment mechanism, may explain the depolarization effect. The turbulent structure of the magnetic field, which can induce the field’s tangling and therefore reduce $P$, also provides a plausible explanation to the decreasing of the polarization percentage towards higher intensities \citep{2015A&A...576A.105P, 2018arXiv180706212P}. 

We fitted the $P$-$I$ relation with a power-law slope, and found the slope index is $-$1.03 $\pm$ 0.05, which indicates that the polarized intensity is almost constant in Oph-C. The nearly constant polarized intensity is more clearly shown in Figure~\ref{fig:figpi}(b). The slope index for Oph-C is slightly lower than the index of $-$0.92 $\pm$ 0.05 for the entire Ophiuchus cloud \citep{2015A&A...576A.105P}. For other dense cores in Ophiuchus, a slope index of -0.7 $\sim$ -0.8 was found in Oph-A \citep{2018ApJ...859....4K}, and an index of around -0.9 was found in Oph-B \citep{2018ApJ...861...65S}. Considering that Oph-A is the warmest and most evolved among the Oph cores, Oph-B is more quiescent than Oph-A, and Oph-C is the most quiescent region in Ophiuchus \citep{2015MNRAS.450.1094P}, it appears that the power-law slope of the $P$-$I$ relation is shallower in more evolved dense cores in the Ophiuchus cloud. This trend could be explained by the improved alignment efficiency resulting from the additional internal radiation \citep[predicted by the RAT theory][]{2007JQSRT.106..225L} in more evolved regions. Alternatively, if the depolarization is caused by turbulence \citep{2015A&A...576A.105P, 2018arXiv180706212P}, the stronger turbulence in more evolved dense cores \citep{2007A&A...472..519A} may also be a possible reason for the variation in the slope index. More detailed analysis of the depolarization effect in the Ophiuchus Cloud will be presented in a separate publication by the BISTRO team. 

\subsection{Magnetic field strength}

\subsubsection{Comparison of three modified DCF methods}\label{methods}

While the morphologies of magnetic fields can help us to qualitatively understand its role in the star formation process, the magnetic field strength is important in quantitatively assessing the significance of the magnetic field compared to gravity based on the mass-to-flux ratio, and compared to turbulence based on the ratio of random-to-ordered components in polarization angle statistics. The strengths of magnetic fields, however, cannot be measured directly from polarization observations. In this paper, we estimated the average magnetic field strength in Oph-C from different modified DCF methods. Results of these methods are shown in Table \ref{tab:para_int} and Table \ref{tab:para}. From the statistical analyses of the dispersion of dust polarization angles, the beam-integrated angular dispersions derived from the SF and ACF methods are consistent with each other ($\sim$21$\degr$ to $\sim$22$\degr$), and are larger than that derived from the UM method ($\sim$11$\degr$), indicating the magnetic field strength estimated from the UM method could be systematically larger than that derived from the SF method and the ACF method. Similar behavior was found when applying these methods on the polarization map of OMC-1 \citep{2009ApJ...696..567H, 2009ApJ...706.1504H, 2017ApJ...846..122P}, a region that has a relatively stronger magnetic field (the $B_{\rm pos}$ is $\sim$13.2 mG estimated from the UM method and $\sim$3.5 to $\sim$3.8 mG estimated from the SF/ACF method without correction for beam integration) than that in Oph-C. The estimated $B_{\rm pos}$ in Oph-C ($\sim$0.1 to $\sim$0.2 mG) is lower than $B_{\rm pos}$ in Oph-A ($\sim$0.2 to $\sim$5 mG) and Oph-B ($\sim$0.6 mG).

Because the results of the dispersion function analysis could be affected by the bin size \citep{2010ApJ...721..815K}, we have redone the SF and ACF analyses to find the dependence on the bin size. We found that oversampling (with bin size $<$7$\arcsec$) would inject additional noises into the dispersion functions (both SF and ACF), thus leading to overestimation of the angular dispersion and underestimation of the B-field strength. The origin of the additional noise is possibly related to the wrongly generated masks due to small pixel size in the POL-2 data reduction process, and needs to be further investigated. Increasing the bin size, on the other hand, shows little effects on the SF method, and leads to larger values of turbulent scale and B-field strength for the results of the ACF method. We also found that, by undersampling, the turbulent scale estimated from the ACF method is always approximately equal to the bin size. The effects on the ACF method can be simply explained by a loss of information on small scales due to undersampling. \citet{2010ApJ...721..815K} has also investigated the dependence of the SF method on the bin size, but got different results from ours: oversampling shows little effect on the SF, while undersampling biases the analysis toward larger dispersion values. This indicates that the dependence of dispersion function on the bin size is not simple. Considering these factors and that the derived turbulent scale ($\sim$ 7.0$\arcsec$) is approximately equal to the Nyquist sampling interval of our data, we note that the turbulent scale along with the B-field strength derived from the ACF method could be overestimated.

Increasing the box size for smoothing would significantly overestimate the angular dispersion derived from the UM method because of field curvature, while in a zero-curvature case, decreasing the box size would slightly underestimate the angular dispersion \citet{2017ApJ...846..122P}. Since the B-field in Oph-C does not show well-defined shapes, it is unclear whether the angular dispersion in Oph-C is underestimated or overestimated by the UM method. We checked the dependence on the box size in our UM analysis by re-applying the UM method to our data with a 5 $\times$ 5 pixel smoothing box and a 7 $\times$ 7 pixel smoothing box, and derived angular dispersions of $\sim$11$\degr$ to $\sim$12$\degr$, indicating that larger smoothing box would not significantly change the results of our UM analysis.  

Systematic uncertainties of the DCF method may arise from the beam-integration effect. For the UM method and the SF method, we use a correction factor $Q_c$ to account for the averaging effect of turbulent cells along the line of sight. \citet{2001ApJ...546..980O} found that $Q_c$ is in the range of 0.46-0.51 for angular dispersions less than 25$\degr$. In our case, the Gaussian fitting of the position angles of polarization segments in Oph-C gives a standard deviation of angle of 33$\degr$, which is larger than the angular dispersion limit of \citet{2001ApJ...546..980O}. However, this standard deviation includes the contribution from the curvature of the large-scale field. Excluding the angular variations of the large-scale field, we got standard deviations $<$ 25$\degr$ from the modified DCF methods. So we adopted a conventional $Q_c$ value of 0.5. The uncertainty of the $Q_c$ value is $\sim$30 \% \citep{2004ApJ...600..279C}. On the other hand, the ACF method takes into account the beam-integration effect by directly fitting the angular dispersion function. The number of turbulent cells of $\sim$2.5 derived from the ACF method is equivalent to a $Q_c$ of $\sim$0.63, which is slightly larger than the correlation factor adopted by the SF analysis and the UM analysis.

As mentioned by \citet{2012ARA&A..50...29C}, even applying the most complicated modified DCF method on the highest quality data would lead to a $B_{\rm pos}$ value with an uncertainty varying by a factor of two or more due to various reasons. It is essential to assess the accuracy of these methods by comparing the results of these methods on polarization maps from simulations. Although the magnetic field strengths estimated from the three modified DCF methods may have systematic differences, they are consistent with each other within the uncertainties, indicating that these results are robust to some extent, and that we can still compare the relative importance of magnetic field with gravity and turbulence with these results. 

\subsubsection{Magnetic field vs. gravity}

To find out whether or not the magnetic field can support Oph-C against gravity, we compared the mass-to-magnetic-flux ratio with the critical ratio using the local magnetic stability critical parameter $\lambda$ \citep{2004ApJ...600..279C}: 
\begin{equation}
\lambda = \frac{(M/\Phi)_{observed}}{(M/\Phi)_{critical}},
\end{equation}
where $(M/\Phi)_{observed}$ is the observed mass-to-magnetic-flux ratio: 
\begin{equation}
(\frac{M}{\Phi})_{observed} = \frac{\mu m_{\mathrm{H}}N(H_2)}{B},
\end{equation}
and $(M/\Phi)_{critical}$ is the critical mass-to-magnetic-flux ratio:
\begin{equation}
(\frac{M}{\Phi})_{critical} = \frac{1}{2\pi\sqrt{G}}.
\end{equation}
We estimated $\lambda$ using the relation in \citet{2004ApJ...600..279C}:
\begin{equation}
\lambda = 7.6 \times 10^{-21} \frac{N(H_2)}{B_{\rm pos}}
\end{equation}

The observed critical parameters derived from the SF, ACF, and UM methods are 7.8 $\pm$ 5.7, 5.9 $\pm$ 4.6, and 3.8 $\pm$ 3.0 (see Table \ref{tab:para}), respectively. \citet{2004ApJ...600..279C} proposed that the observed $M/\phi$ along with $\lambda$ are overestimated because of geometrical effects. \citet{1993ApJ...407..175C} found a average line-of-sight B-field strength ($B_{\mathrm{los}}$) of $+$ 6.8 $\pm$ 2.5 $\mu$G in the Ophiuchus cloud based on OH Zeeman observations with a 18$\arcmin$ beam. Their estimated $B_{\mathrm{los}}$ is much smaller than the B$_{\mathrm{pos}}$ infered by our analyses, indicating that the B-field in Oph-C is possibly lying near the plane of sky. However, since quasi-thermal OH emissions cannot trace high density materials with $n$(H$_2$) $> 10^4 $ cm$^{-3}$ and the beam of the OH Zeeman observation is much larger than that of our polarization maps, it is more likely that the line-of-sight B-field strength in the Oph-C is underestimated by \citet{1993ApJ...407..175C}. As the degree of the underestimation of $B_{\mathrm{los}}$ is unknown, the correction factor for the geometrical bias cannot be derived by simply comparing $B_{\mathrm{los}}$ and $B_{\mathrm{pos}}$. Alternatively, we adopted a statistical correction factor of 3 \citep{2004ApJ...600..279C}. By applying this correction, we obtain corrected critical parameters ($\lambda_c$) of 2.6 $\pm$ 1.9, 1.9 $\pm$ 1.5, and 1.3 $\pm$ 1.0 (see Table \ref{tab:para}) for the SF, ACF, and UM methods, respectively. These values indicate that the Oph-C region is near magnetically critical or slightly magnetically supercritical (i.e. unstable to collapse).

\subsubsection{Magnetic field vs. turbulence}

To compare the relative importance of the magnetic field and turbulence in Oph-C, we calculated the magnetic field energy and the internal nonthermal kinetic energy.  The total magnetic field energy is given by: 
\begin{equation}
E_B = \frac{B^2 V}{2 \mu_0}
\end{equation}
in SI units, where $\mu_0$ is the permeability of vacuum and $B = \frac{4}{\pi}B_{\rm pos}$ \citep{2004ApJ...600..279C} is the total magnetic field strength. And the internal nonthermal kinetic energy is derived by: 
\begin{equation}
E_{K,NT} = \frac{3 M \sigma_v^2}{2}.
\end{equation}
For the estimated volume (see Section \ref{sec:strength}) of Oph-C, the internal nonthermal kinetic energy is  (6.1 $\pm$ 2.0) $\times$ 10$^{35}$ J. The total magnetic field energy measured from the SF, ACF, and UM methods are (5.4 $\pm$ 4.8) $\times$ 10$^{35}$ J, (9.5 $\pm$ 9.7) $\times$ 10$^{35}$ J, and (2.3 $\pm$ 2.5) $\times$ 10$^{36}$ J (see Table \ref{tab:para}), respectively. The $E_B$ calculated from the SF method is comparable to $E_{K,NT}$, while the values of $E_B$ estimated from the ACF and UM methods are greater than $E_{K,NT}$. However, the uncertainty is more than 100\% for the values of $E_B$ calculated from the ACF and UM methods. Thus, we can only set upper limits for the total magnetic field energy in Oph-C from these two methods. 

\section{Summary}
As part of the BISTRO survey, we have presented the 850 $\mu$m polarization observations toward the Oph-C region with the POL-2 instrument at the JCMT. The main conclusions of this work are as follows:
\begin{enumerate}
\item Our POL-2 observations are much more sensitive and trace a larger area than previous SCUPOL observations. Unlike the randomly distributed magnetic field orientations traced by the SCUPOL observations, the magnetic field traced by our POL-2 observations show an ordered field geometry with a predominant orientation of northeast-southwest. We found the average angular difference of spatially overlapping vectors between the two data sets to be $\sim$39$\degr$. We performed a K-S test on the position angles, and found that the POL-2 data and the SCUPOL data have low probability (0.06) to be drawn from the same distribution. The inconsistency between the POL-2 and the SCUPOL data may be explained by the low signal-to-noise ratio of the SCUPOL data.
\item The B-field orientation in Oph-C is consistent with the B-field orientations in Oph-A and Oph-B. The orientation also agrees with the B-field component in lower-density regions traced by NIR observations, and is aligned with the cloud-scale B-field orientation revealed by Planck.
\item We detect a decreasing polarization percentage as a function of increasing total intensity in the Oph-C region. The power-law slope index is found to be $-$1.03 $\pm$ 0.05, suggesting that the polarized intensity is almost constant in Oph-C.
\item We compare the plane-of-sky magnetic field strength in Oph-C calculated from different modified DCF methods. The $B_{pos}$ calculated by the SF method, the ACF method, and the UM method are 103 $\pm$ 46 $\mu$G, 136 $\pm$ 69 $\mu$G, and 213 $\pm$ 115 $\mu$G, respectively. 
\item The mass-to-magnetic-flux ratio of Oph-C is found to be comparable to or slightly higher than its critical value, suggesting that the Oph-C region is near magnetically critical magnetically supercritical (i.e., unstable to collapse).
\item In Oph-C, the total magnetic energy calculated from the SF method is comparable to the turbulent energy. Due to large uncertainties, the ACF method and the UM method only set upper limits for the total magnetic energy.
\item We compared our work with studies of two other dense cores in the Ophiuchus cloud. We find the B-fields in Oph-C and Oph-B have larger angular dispersions than Oph-A. We also find a possible trend of shallower $P$-$I$ relationship with evolution in the three dense cores in the Ophiuchus region. In addition, the $B_{pos}$ in Oph-C is lower than $B_{pos}$ of more evolved regions (e.g., Oph-A and Oph-B) in Ophiuchus.

\end{enumerate}

\acknowledgments  We thank Dr. Qizhou Zhang for helpful discussions on dispersion function analysis. J.L., K.Q., D.L., and L.Q. are supported by National Key R\&D Program of China No. 2017YFA0402600. J.L. and K.Q. acknowledge the support from National Natural Science Foundation of China (NSFC) through grants U1731237, 11590781, and 11629302. J.L. acknowledges the support from the program of China Scholarship Council (No. 201806190134) and from the Smithsonian Astrophysical Observatory pre-doctoral fellowship. K.P. was an International Research Fellow of the Japan Society for the Promotion of Science. D.W.T. and K.P. acknowledge Science and Technology Facilities Council (STFC) support under grant numbers ST/K002023/1 and ST/M000877/1. J.K. was supported by MEXT KAKENHI grant number 16H07479 and the Astrobiology Center of NINS. M.T. was supported by MEXT KAKENHI grant number 22000005. D.W.T. and K.P. acknowledge Science and Technology Facilities Council (STFC) support under grant numbers ST/K002023/1 and ST/M000877/1. C.W.L. and M.K. were supported by the Basic Science Research Program through the National Research Foundation of Korea (NRF), funded by the Ministry of Education, Science and Technology (CWL: NRF-2016R1A2B4012593) and the Ministry of Science, ICT and Future Planning (MK: NRF-2015R1C1A1A01052160). K.P. and S.P.L. acknowledge the support of the Ministry of Science and Technology of Taiwan (Grant No. 106-2119-M-007-021-MY3). W.K. was supported by Basic Science Research Program through the National Research Foundation of Korea (NRF-2016R1C1B2013642). J.E.L. is supported by the Basic Science Research Program through the National Research Foundation of Korea (grant No. NRF-2018R1A2B6003423) and the Korea Astronomy and Space Science Institute under the R\&D program supervised by the Ministry of Science, ICT and Future Planning. A.S. acknowledge the support from KASI for postdoctoral fellowship. T.L. is supported by a KASI fellowship and an EACOA fellowship. The James Clerk Maxwell Telescope is operated by the East Asian Observatory on behalf of the National Astronomical Observatory of Japan, the Academia Sinica Institute of Astronomy and Astrophysics, the Korea Astronomy and Space Science Institute, the National Astronomical Observatories of China, and the Chinese Academy of Sciences (Grant No. XDB09000000), with additional funding support from the Science and Technology Facilities Council of the United Kingdom and participating universities in the United Kingdom and Canada. The James Clerk Maxwell Telescope has historically been operated by the Joint Astronomy Centre on behalf of the Science and Technology Facilities Council of the United Kingdom, the National Research Council of Canada, and the Netherlands Organisation for Scientific Research. Additional funds for the construction of SCUBA-2 and POL-2 were provided by the Canada Foundation for Innovation. The data used in this paper were taken under project code M16AL004. This research made use of APLpy, an open-source plotting package for Python \citep{2012ascl.soft08017R}, Astropy, a community-developed core Python package for Astronomy \citep{2013A&A...558A..33A}, and Matplotlib, a Python 2D plotting library for Python \citep{2007CSE.....9...90H}.

\facility{James Clerk Maxwell Telescope (JCMT).}

\software{APLpy \citep{2012ascl.soft08017R}, Astropy \citep{2013A&A...558A..33A},
          Matplotlib \citep{2007CSE.....9...90H}, Starlink \citep{2005ASPC..343...71B, 2013StaUN.258.....C}}

\appendix
\section{Uncertainty from sparse sampling}\label{modeling}
Here we derive the uncertainty in the dispersion function caused by the lack of vector samples (sparse sampling). We perform simple Monte Carlo simulations of modeled structured fields with randomly generated Gaussian dispersions to roughly estimate the uncertainty of sparse sampling in the dispersion function of our data. It should be noted that because simulating the beam-integration effect is extremely time-consuming and only affects the first two or three data points of the dispersion function, the beam-integration effect is not taken into account in our toy models. 

\begin{figure*}[!tbp]
 \gridline{\fig{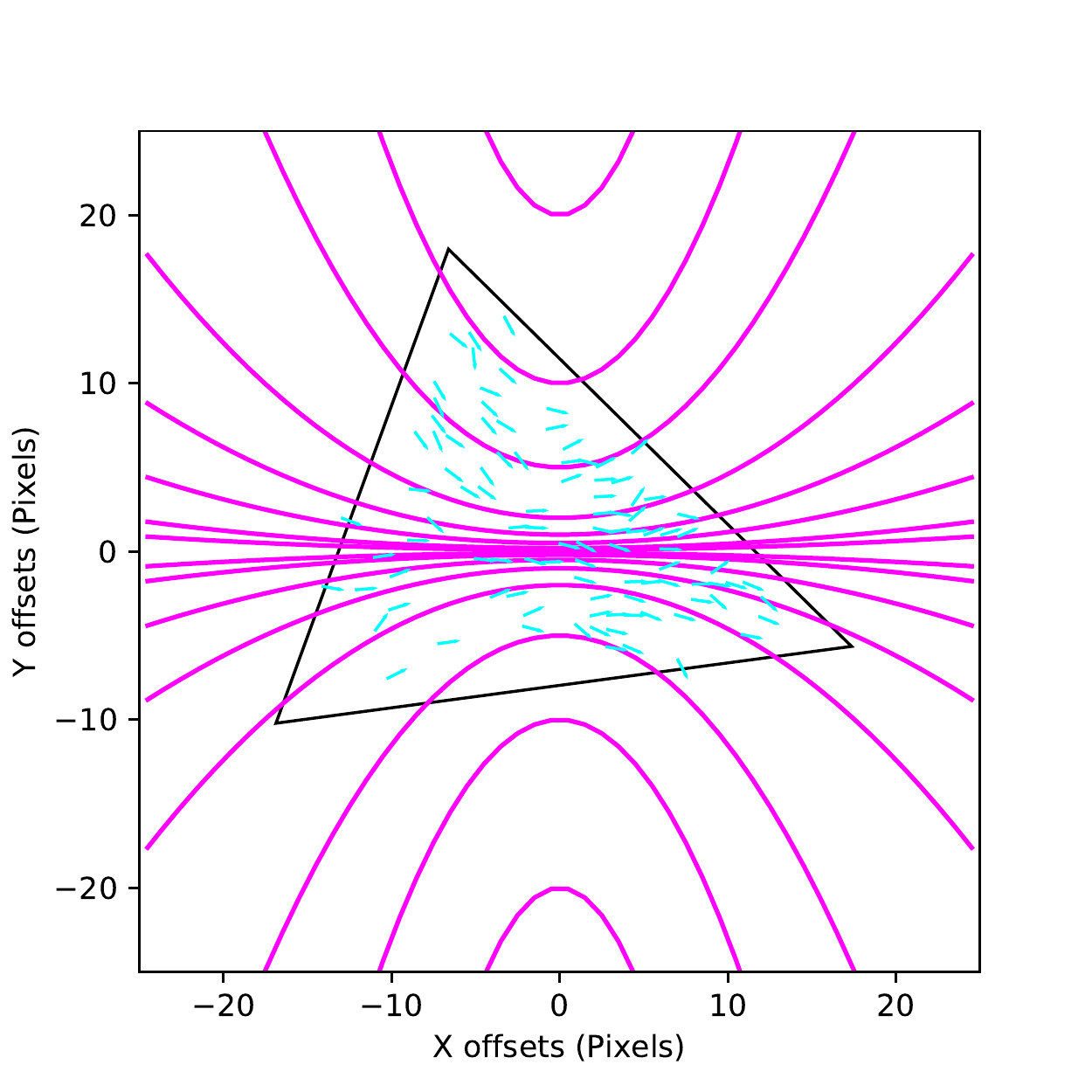}{0.5\textwidth}{(a)}
      \fig{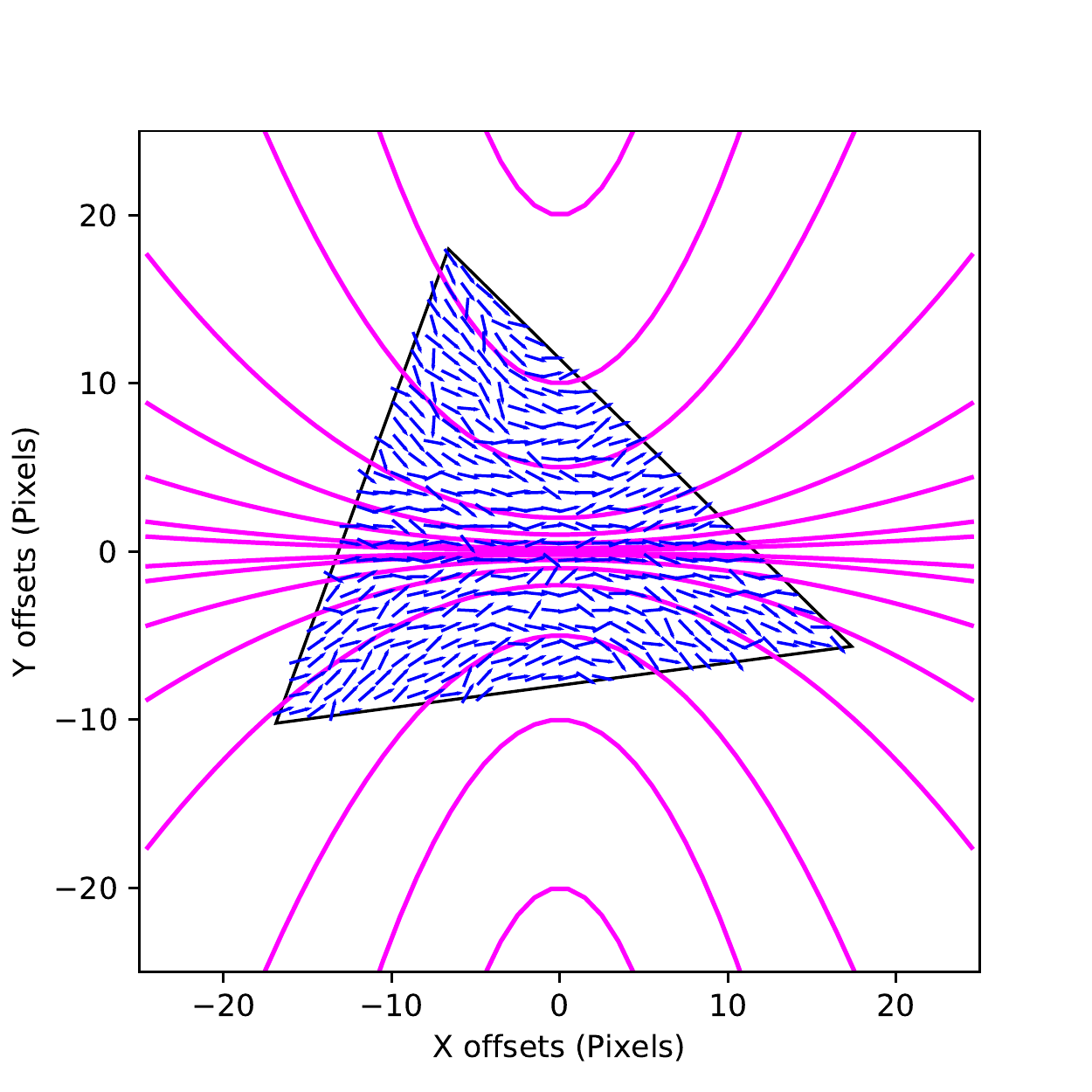}{0.5\textwidth}{(b)}}
\gridline{\fig{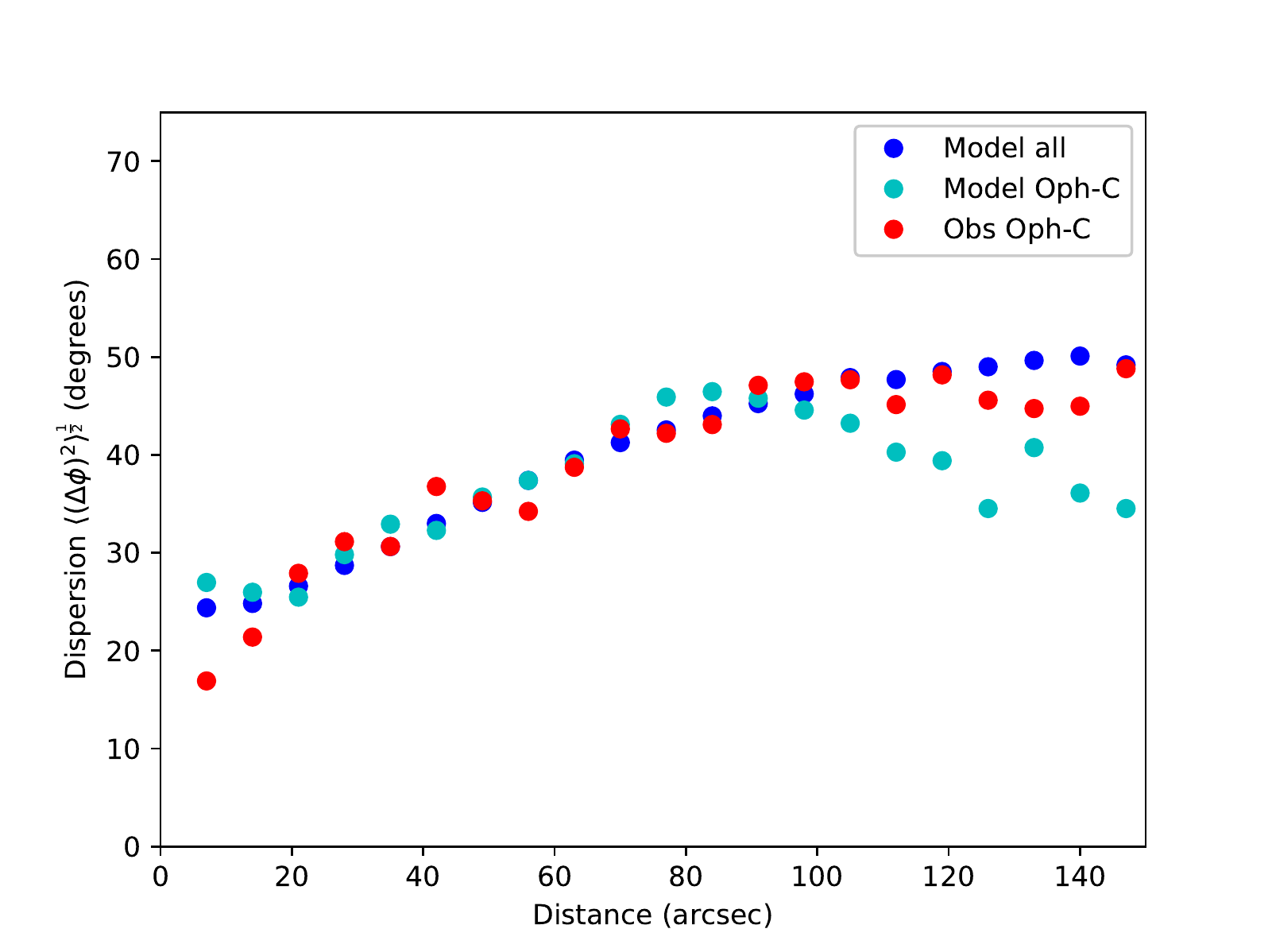}{0.5\textwidth}{(c)}
        \fig{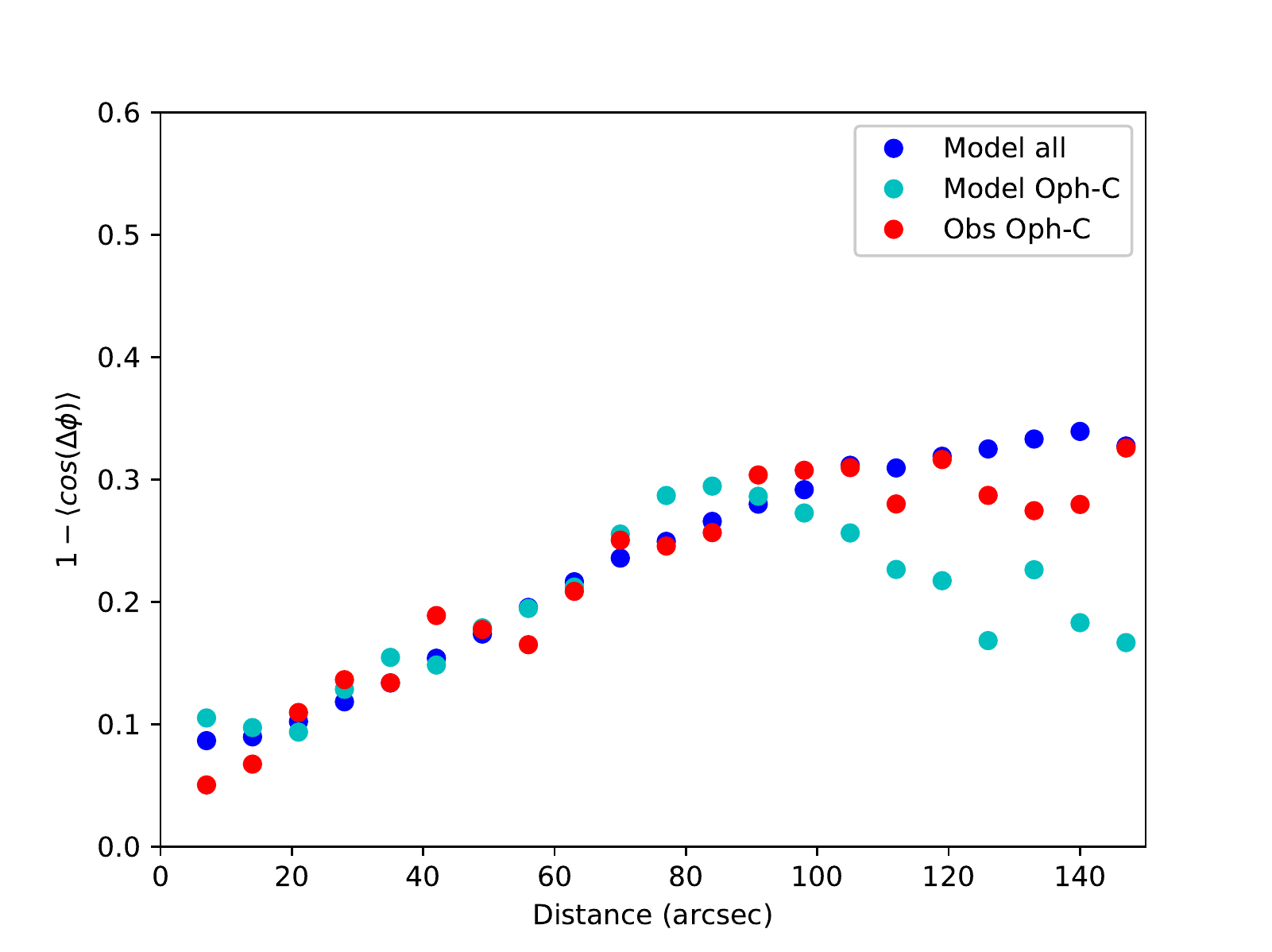}{0.5\textwidth}{(d)}}

\caption{(a) The modeled sparsely sampled B-vectors with an angular dispersion of 22$\degr$ are shown in cyan. (b) The modeled ``unbiased'' B-vectors with an angular dispersion of 22$\degr$ are shown in blue. In panel (a) and (b), magenta curves denote the underlying parabola field models with $C = 0.13 $. Vectors are of unit length. And the triangle marks the region in which we calculated the ``unbiased'' dispersion function. (c) Structure functions calculated from samples corresponding to all modeled vectors in the triangle region (blue), modeled Oph-C vectors (cyan), and observed Oph-C vectors (red). (d) Auto-correlation function with the same symbols as those in panel (c).  \label{fig:figmod}}
\end{figure*}

We start with generating the underlying field model. We note that since the uncertainty in the angular dispersion function due to sparse sampling is only related to the amount of spatial correlation of field orientations across the sky and the amount of angular dispersion relative to the structured field \citep{2016A&A...596A..93S}, the choice of the underlying field model is arbitrary. We build a set of underlyling parabola models \citep[e.g., ][]{2006Sci...313..812G, 2009ApJ...707..921R, 2014ApJ...794L..18Q} with the form: 
\begin{equation}
y = g + gCx^2,
\end{equation}
where x is the offsets in pixels along the field axis from the center of symmetry. In Figure \ref{fig:figmod} (a) and (b), a parabola field model with $C = 0.13$ is shown in magenta curves as an example.

We then derive the orientation of the modeled sparsely sampled B-vectors by applying a Gaussian angular dispersion of 22$\degr$ (to match the angular dispersion of $\sim$21$\degr$ to $\sim$22$\degr$ derived from the SF and ACF methods) to the underlying B-vectors with the same spatial distributions as those of the observed B-vectors in Oph-C with $P/\delta P>3$ and $\delta P<5$\%, while the offsets and angle of the modeled B-vectors with respect to the center of symmetry of the underlying modeled field is random. An example of the modeled sparsely sampled B-vectors is shown in Figure \ref{fig:figmod} (a).

In a similar way, we also derived the orientation of  ``unbiased'' samples of B-vectors with a Gaussian angular dispersion of 22$\degr$ and spatial separation of 1 pixel (7$\arcsec$) for comparison. There are enough vectors in the ``unbiased'' sample to achieve statistical significance. An example of the modeled ``unbiased'' B-vectors is shown in Figure \ref{fig:figmod} (b).

We calculate the SF and ACF (see Figure \ref{fig:figmod} (c) and (d) for examples of the SF and ACF) from the sparse samples and ``unbiased'' samples of modeled vectors, and find that the average deviation of the SFs and ACFs between the two sets of samples are $\sim$1.5$\degr$ and $\sim$0.015 over $25\arcsec<l<100\arcsec$, relatively for SFs and ACFs with similar amounts of large-scale spatial correlation and random anglular dispersions (e.g., similar SF and ACF shapes over $25\arcsec<l<100\arcsec$) to the dispersion functions calculated from the observed data. These average deviations, which are larger than the statistical uncertainties ( $\sim$0.6$\degr$ for the SF and $\sim$0.007 for the ACF over $25\arcsec<l<100\arcsec$ in average) propagated from the measurement uncertainty, are introduced in our analyses as the uncertainties due to sparse sampling. 

\end{document}